\documentclass[preprint]{aastex631}
\usepackage{amsmath}

\shorttitle{ELLIPTICITY ANGLE STATISTICS}
\shortauthors{McKINNON}
\begin{document}

\title{Statistical Properties of a Polarization Vector's Ellipticity Angle}
\author{M. M. McKinnon}
\affiliation{National Radio Astronomy Observatory, Socorro, NM \ 87801 \ \ USA}
\correspondingauthor{M. M. McKinnon}
\email{mmckinno@nrao.edu}


\begin{abstract}
The orientation of a polarization vector on the Poincar\'e sphere is defined by its position
angle (PA) and ellipticity angle (EA). The radio emission from pulsars, magnetars, and 
fast radio bursts can be elliptically polarized, and measurements of the EA have become 
increasingly important in interpretations and models of their polarization. An in-depth
understanding of the statistical properties of the measured polarization angles is a 
prerequisite to their detailed interpretation. While the statistics of the PA have been 
understood for some time, the statistics of the EA do not appear to be as well developed 
as those of the PA. The statistical properties of the EA are derived when the amplitude 
of the polarization vector is constant, to include its probability density, mean, standard 
deviation, and confidence limits. Similar to the PA, the standard deviation and confidence 
limits of the EA vary inversely with the polarization signal-to-noise ratio. However, 
unlike the PA, the probability density of the EA is generally asymmetric, its standard 
deviation and confidence limits are dependent upon the intrinsic value of the EA, and the 
measured EA is biased by the instrumental noise, particularly at low signal-to-noise ratios 
and large values of the intrinsic EA. General expressions for the joint probability density 
of the polarization angles and the probability density of the EA are also derived when the 
amplitude of the polarization vector fluctuates due to the superposition of incoherent modes 
of orthogonal polarization.
\end{abstract}


\section{INTRODUCTION}

Pulsar average profiles have typically shown how the total intensity, circular polarization, linear 
polarization, and polarization position angle (PA) vary across the pulsar's pulse. In retrospect,
measurements of the polarization vector's ellipticity angle (EA) have been absent from most of 
these profiles. Both the PA and the EA are needed to completely define the orientation of the 
vector on the Poincar\'e sphere. The observations of PSR B0329+54 by Edwards \& Stappers (2004) and 
PSR B0809+74 by Edwards (2004) were some of the first to investigate the behavior of the EA in detail. 
At some pulse longitudes in both pulsars, individual samples of the EA are concentrated near the 
equator of the Poincar\'e sphere, but the average EA makes a large excursion toward one of the 
poles of the sphere. These excursions tend to be accompanied by a large discontinuity in the PA of 
$\Delta\psi\simeq\pi/2$. The PA discontinuities are indicative of orthogonal modes of polarization
(OPMs), which are generally attributed to the natural modes of wave propagation in the pulsar's
magnetosphere (Melrose \& Stoneham 1977; Melrose 1979; Allen \& Melrose 1982; Barnard \& Arons 1986). 
The PA and EA in parts of PSR B0031-07 (Ilie et al. 2020) behave similarly to those of PSRs B0329+54 
and B0809+74. The PA and EA in the central component of PSR B0329+54 combine to form an intriguing 
partial annulus in one hemisphere of the sphere. Edwards \& Stappers (2004) suggested the annulus 
may be the result of generalized Faraday rotation in the pulsar magnetosphere (Kennett \& Melrose 
1998). They attributed the large EA excursions observed elsewhere in the pulse to either a transition 
between polarization modes that were not precisely orthogonal or an OPM transition combined with a 
circularly polarized emission component.

More recent observations of pulsars, magnetars, and fast radio bursts (FRBs) have revealed 
large variations in the EA as functions of pulse longitude, frequency, or time. In some cases, 
the trajectories of their polarization vectors trace arcs or great circles on the Poincar\'e 
sphere. The magnitude of the EA can be large as the vector approaches a pole of the sphere. 
Dyks et al. (2021) attributed the longitude-dependent variations in the EA of PSR B1451-68 to 
changes in the relative intensity of coherent OPMs. They recognized that the variations in 
PA and EA caused by such an OPM transition follow a great circle on the Poincar\'e sphere, whereas 
variations due to changes in the phase offset between the modes generally trace a small circle on 
the sphere. Oswald et al. (2023) proposed that the OPMs in a pulsar's magnetosphere are partially 
coherent and demonstrated that the frequency-dependent variations in the EA of PSR J0820-1350 
and the longitude-dependent EA variations in PSR J1157-6224 are consistent with changes in the 
phase offset between the modes. They also demonstrated that the longitude-dependent variations 
in the EA of PSR J0134-2937 are consistent with a change in the relative intensity of the modes. 
Lower et al. (2024) found that frequency-dependent variations in the PA and EA of the magnetar 
XTE J1810-197 follow a circle on the Poincar\'e sphere. The possible explanations for this behavior 
include Faraday conversion in the relativistic magnetized plasma that surrounds the magnetar or 
mode coupling within its magnetosphere. Bera et al. (2025) showed that temporal variations in the 
PA and EA of FRBs 20210912A and 20230708A follow great circles on the sphere. The variations are 
consistent with propagation effects in a birefringent medium located in the outer magnetosphere 
of the FRB progenitor or in the magnetized plasma surrounding it. Cao et al. (2025) observed
large variations in the PA and EA across single pulses of PSR B1919+21. They attributed the 
variations to changes in the phase offset of partially coherent polarization modes. The 
aforementioned examples illustrate a variety of EA behaviors and emphasize the importance of 
the EA in the interpretation of the observations, particularly within the context of different 
polarization models.

An in-depth understanding of the statistical properties of the polarization angles is a
prerequisite to their detailed interpretation. The statistical properties of the PA have been 
understood for quite some time (e.g., Davenport \& Root 1958; Papoulis 1965; Naghizadeh-Khouei 
\& Clarke 1993). The probability density function (pdf) of the PA, $f_\psi(\psi)$, is unimodal 
and symmetric about its intrinsic value, $\psi_o$. When the signal-to-noise ratio (SNR) in 
linear polarization is large ($s\gg 1)$, $f_\psi(\psi)$ is Gaussian, with a mean of $\psi_o$ 
and a standard deviation of $\sigma_\psi=1/(2s)$. The PA pdf is uniform when $s=0$. It is 
straightforward to numerically derive the confidence limits of $f_\psi(\psi)$, because it is 
symmetric about $\psi_o$ (Wardle \& Kronberg 1974; Naghizadeh-Khouei \& Clarke 1993; Everett 
\& Weisberg 2001). And since the pdf is Gaussian at high SNR, the confidence limits can be 
approximated by its standard deviation. McKinnon \& Stinebring (1998) generalized $f_\psi(\psi)$ 
to include OPMs. This pdf is generally bimodal. The angular separation between the two peaks of 
the pdf is $\Delta\psi=\pi/2$, the difference between the heights of the peaks is proportional 
to the difference in the mean intensities of the modes, and the widths of the peaks vary 
inversely with the polarization fluctuations (McKinnon 2022).

The statistical properties of the EA do not appear to be as well developed as those for the PA. 
McKinnon (2003, 2006) derived the pdf of the polarization vector's colatitude, $\theta$, which 
is complementary to twice the EA ($\theta=\pi/2-2\chi$). The derivation was limited to circularly 
polarized radiation. For this specific case, the pdf of $\theta$ evolves from a sine distribution, 
$f_\theta(\theta)=\sin(\theta)/2$, when the SNR in total polarization is $s=0$ to a Rayleigh 
distribution when $s\gg 1$. The EA pdf is also bimodal when the OPMs are circularly polarized.

The purpose of this paper is to derive the general statistical properties of the EA, to 
include its pdf, mean, standard deviation, and confidence limits. The paper is organized as 
follows. A general expression for the joint probability density of the PA and EA is derived 
in Section~\ref{sec:fix} for the case when the amplitude of the polarization vector is constant. 
The pdf of the EA is derived from the joint density. The mean and standard deviation of the EA 
are then calculated from its pdf. The EA statistics derived in this section are complementary 
to the PA statistics documented in Naghizadeh-Khouei \& Clarke (1993). In Section~\ref{sec:opm}, 
the joint density and EA pdfs are derived for the case when the amplitude of the polarization 
vector fluctuates due to the superposition of incoherent OPMs. The EA statistics developed in 
this section are complementary to those of the PA derived for OPMs in McKinnon \& Stinebring 
(1998). The results of the analysis are compared with observations in Section~\ref{sec:obs}. 
The statistical properties of the PA and EA are compared and contrasted in 
Section~\ref{sec:compare}. Summary comments are listed in Section~\ref{sec:summary}. A method 
for calculating the confidence limits of the EA is proposed in the Appendix.


\section{EA of a Fixed Polarization Vector}
\label{sec:fix}

\subsection{Joint Probability Density of the Polarization Angles}
\label{sec:fixjoint}

The procedure for deriving the joint probability density of the polarization angles is outlined 
in McKinnon (2003). To briefly summarize that analysis, the Stokes parameters $Q$, $U$, and $V$ 
are independent Gaussian random variables when the amplitude and orientation of the polarization 
vector are fixed. Their joint probability density is then the product of their individual 
distributions. The joint probability density of the polarization angles can be found by converting 
the Cartesian coordinates of the Stokes parameters to spherical coordinates and integrating over 
radius. For a polarization vector with amplitude $\mu_p$, an intrinsic PA $\psi_o$, and an 
intrinsic EA $\chi_o$, the general expression for the joint probability density of $\psi$ and 
$\chi$ is
\begin{eqnarray}
f(\psi,\chi) & = & \frac{\cos(2\chi)}{\pi}\exp{\left(-\frac{s^2}{2}\right)}
                   \Biggl\{sg(\psi,\chi)\sqrt{\frac{2}{\pi}} 
              + [1+s^2g^2(\psi,\chi)]\exp{\left[\frac{s^2g^2(\psi,\chi)}{2}\right]} \nonumber \\
        & \times & \left[1+{\rm erf}\left(\frac{sg(\psi,\chi)}{\sqrt{2}}\right)\right]\Biggr\}, 
\label{eqn:fixjoint}
\end{eqnarray}
%
%
%
where $s=\mu_p/\sigma_n$ is the SNR in total polarization, $\sigma_n$ is the instrumental noise,
and $g(\psi,\chi)$ is a function that accounts for the polarization vector's geometry. It is 
given by
\begin{equation}
g(\psi,\chi) = \sin(2\chi)\sin(2\chi_o) + \cos(2\chi)\cos(2\chi_o)\cos[2(\psi-\psi_o)].
\end{equation}
To simplify the analysis, and without loss of generality, the intrinsic PA is set to
$\psi_o=0$ in what follows. 

\begin{figure}
\plotone{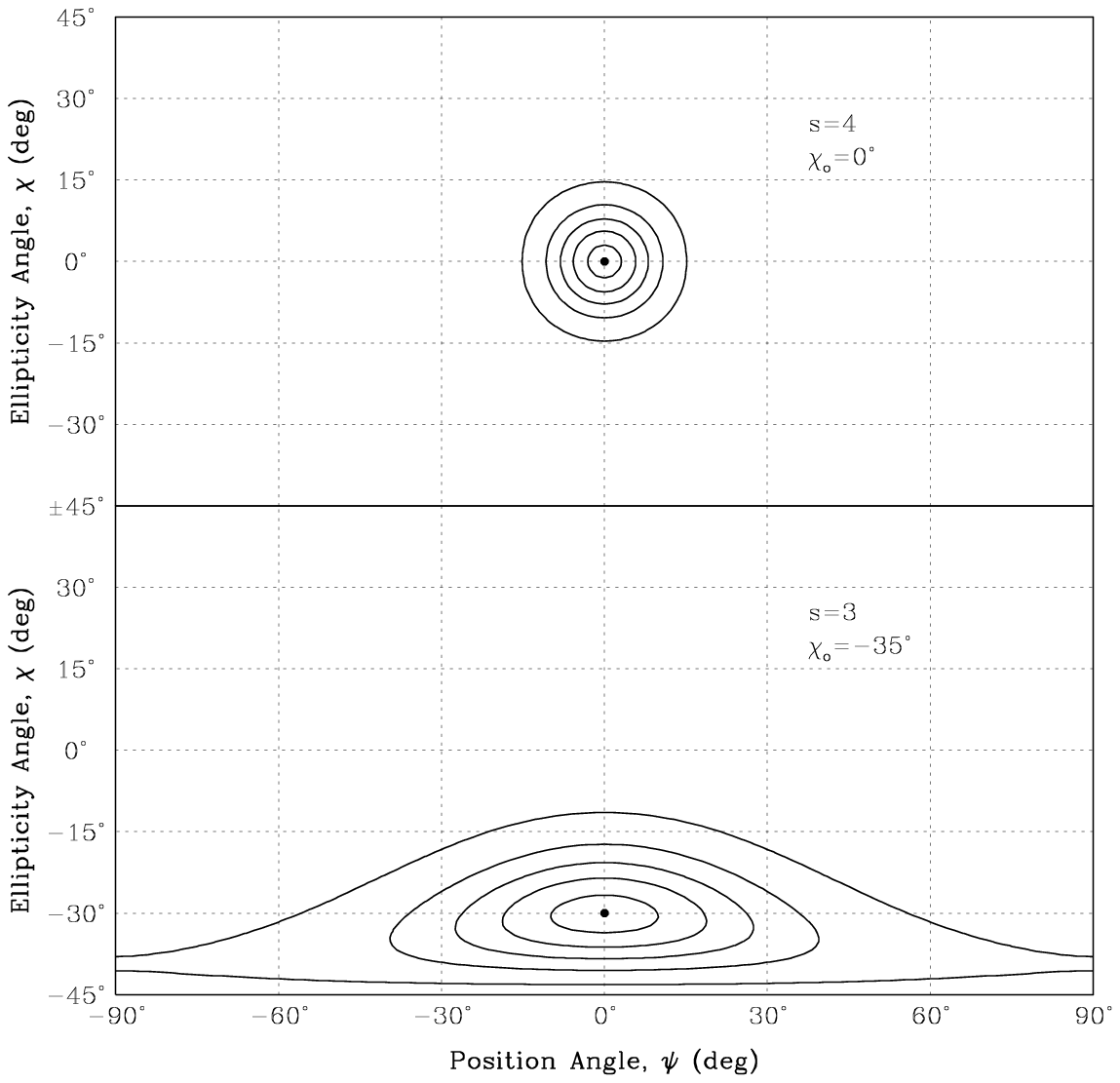}
\caption{Examples of the joint probability density, $f(\psi,\chi)$, of the PA and EA when
the amplitude of the polarization vector is constant (Equation~\ref{eqn:fixjoint}). The 
top panel shows $f(\psi,\chi)$ when $s=4$ and $\chi_o=0^\circ$. The dot marks the peak of 
$f(\psi,\chi)$ at $\psi=\chi=0^\circ$. The bottom panel shows $f(\psi,\chi)$ when $s=3$ 
and $\chi_o=-35^\circ$. The dot marks the peak of $f(\psi,\chi)$ at $\psi=0^\circ$ and 
$\chi=-30^\circ$. The contour levels in both panels are drawn at 0.1, 0.3, 0.5, 0.7, and 
0.9 of the peak value.} 
\label{fig:fixjoint}
\end{figure}

Examples of $f(\psi,\chi)$ are shown in Figure~\ref{fig:fixjoint}. The top panel shows 
$f(\psi,\chi)$ when $s=4$ and $\chi_o=0^\circ$. The radiation is linearly polarized in 
this example, and the joint density is circularly symmetric about its centroid. The 
bottom panel shows $f(\psi,\chi)$ when $s=3$ and $\chi_o=-35^\circ$. The joint density 
is symmetric in $\psi$ but asymmetric in $\chi$. Therefore, the peak of $f(\psi,\chi)$ 
coincides with $\psi_o$ but not with $\chi_o$. Since the radiation is predominantly 
circularly polarized in this example, samples of the PA occur across its entire range. 
When $s=0$, the joint density is isotropic, the PA distribution is uniform, and the EA 
follows a cosine distribution, $f_\chi(\chi)=\cos(2\chi)$.


\subsection{Probability Density of the EA}
\label{sec:chiPDF}

\begin{figure}
\plotone{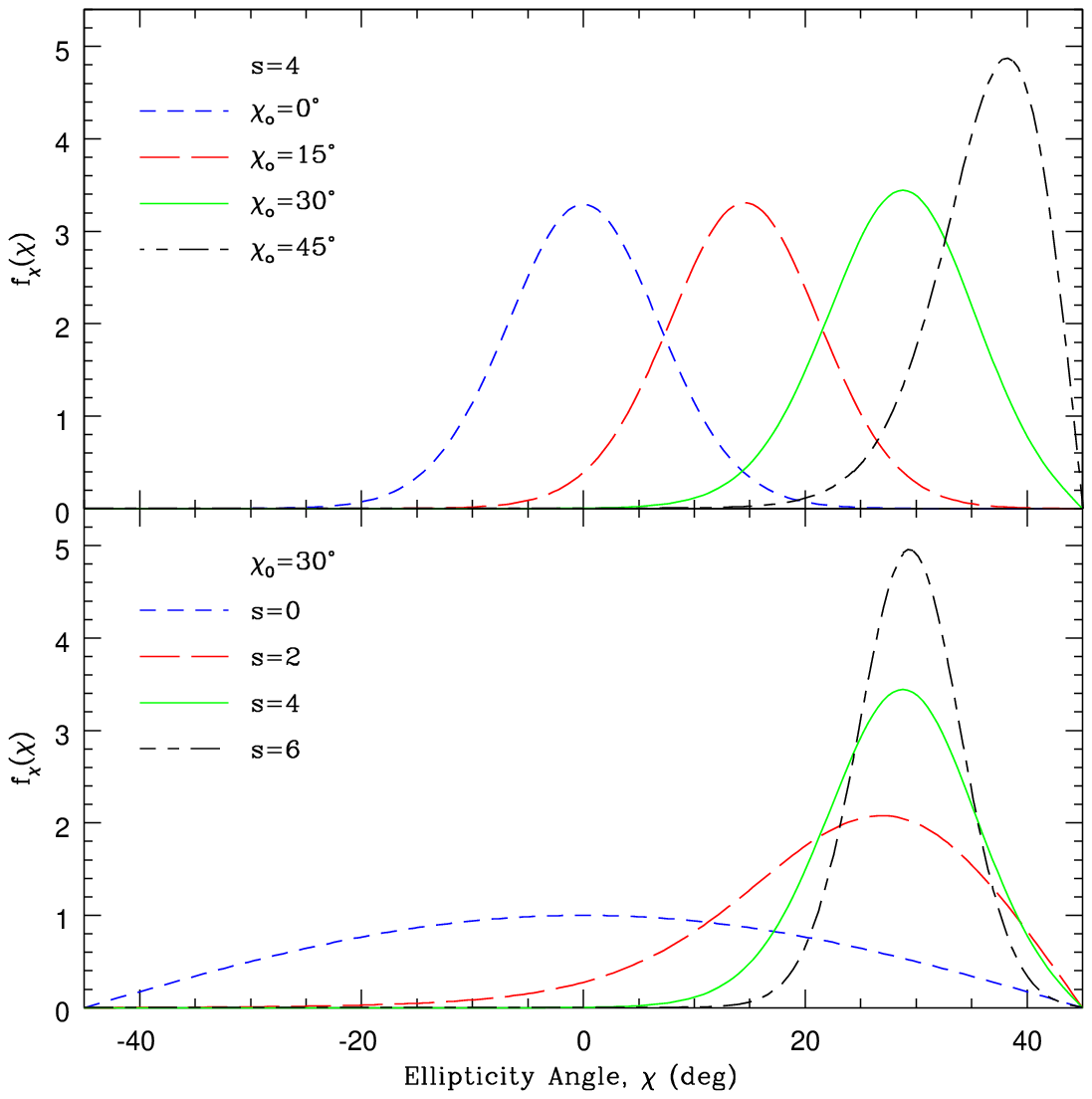}
\caption{Examples of the EA pdf when the amplitude of the polarization vector is constant. 
The top panel shows the pdfs for different values of the intrinsic EA, $\chi_o$, at a 
constant value of the SNR in total polarization, $s=\mu_p/\sigma_n=4$. The bottom panel 
shows the pdfs for different values of $s$ when $\chi_o$ is held constant at $30^\circ$.}
\label{fig:fixpdf}
\end{figure}

The pdf of $\chi$, $f_\chi(\chi)$, can be derived by integrating Equation~\ref{eqn:fixjoint} 
over $\psi$. The resulting pdf is a function of $s$ and $\chi_o$. The pdfs for $\chi_o=0$ and 
$\chi_o=\pi/4$ can be derived analytically. The pdfs for other values of $\chi_o$ can be 
calculated numerically. Examples of $f_\chi(\chi)$ are illustrated in Figure~\ref{fig:fixpdf}. 
The top panel of the figure shows how $f_\chi(\chi)$ evolves as the intrinsic EA changes from 
$\chi_o=0$ to $\chi_o=\pi/4$ at a fixed value of SNR in total polarization. The pdfs are 
symmetric and Gaussian-like for small to intermediate values of $\chi_o$. The pdfs become 
slightly asymmetric as $\chi_o$ increases and evolve to what resembles a mirror image of a 
Rayleigh distribution at $\chi_o=\pi/4$. The bottom panel of the figure demonstrates how the 
pdf evolves with increasing SNR while the intrinsic EA remains constant at $\chi_o=\pi/6$. 
The low to moderate values of $s$ used in the panel were chosen to show that the shape of 
the pdf changes significantly in this SNR regime. The pdf is a cosine distribution when the 
radiation is unpolarized ($s=0$). As the SNR increases, the pdf is initially asymmetric and 
becomes increasingly symmetric, until it resembles a Gaussian distribution at high SNR.

In the case of $\chi_o=0$, $g(\psi,\chi)=\cos(2\chi)\cos(2\psi)$, and the pdf of $\chi$ 
can be derived by first integrating the joint probability density of the Stokes parameters
over $\psi$ and then radius:
\begin{equation}
f_\chi(\chi) = \cos(2\chi)\exp{\left(-\frac{s^2}{2}\right)}{}_1F_1[3/2, 1; s^2\cos^2(2\chi)/2].
\label{eqn:chilin}
\end{equation}
The function ${}_1F_1(a,b;z)$ in Equation~\ref{eqn:chilin} is the confluent hypergeometric function 
with numerical parameters $a$ and $b$ and argument $z$. An example of Equation~\ref{eqn:chilin} 
is shown by the dashed blue line in the top panel of Figure~\ref{fig:fixpdf}. The pdf is symmetric 
about $\chi=0$ and approaches a Gaussian distribution with a standard deviation of 
\begin{equation}
\sigma_\chi\simeq\frac{1}{2s} = \frac{28.65^\circ}{s}
\label{eqn:chiSD0}
\end{equation}
as $s$ becomes large. The dependence of $\sigma_\chi$ upon $s$ in this case is identical to 
that for the standard deviation of the PA, $\sigma_\psi$ (Naghizadeh-Khouei \& Clarke 1993; 
Everett \& Weisberg 2001). Therefore, the joint density of the polarization angles is circularly 
symmetric about $\psi=\chi=0$, as shown in the top panel of Figure~\ref{fig:fixjoint}.

For $\chi_o=\pi/4$, $\psi$ and $\chi$ are independent of one another, and their joint probability 
density is symmetric about the Stokes $V$ axis of the Poincar\'e sphere. The function $g(\psi,\chi)$
is equal to $\sin(2\chi)$, and the pdf of $\chi$ is (see Equation 17 of McKinnon 2003)
\begin{eqnarray}
f_\chi(\chi) & = & \cos(2\chi)\Biggl\{s\sin(2\chi)\sqrt{\frac{2}{\pi}}
                   \exp{\left(-\frac{s^2}{2}\right)}
              + [1+s^2\sin^2(2\chi)]\exp{\left[-\frac{s^2\cos^2(2\chi)}{2}\right]} \nonumber \\
        & \times & \left[1+{\rm erf}\left(\frac{s\sin(2\chi)}{\sqrt{2}}\right)\right]\Biggr\}.
\label{eqn:pdf45}
\end{eqnarray}
%
%
%
An example of Equation~\ref{eqn:pdf45} is shown by the dashed black line in the top panel of
Figure~\ref{fig:fixpdf}. The mean, mode, and standard deviation of this pdf can be estimated 
by recalling that the pdf of the vector colatitude, $f_\theta(\theta)$, approaches a Rayleigh 
distribution at high SNR in total polarization (Equation 18 of McKinnon 2003): 
\begin{equation}
f_\theta(\theta)=s^2\theta\exp{\left(-\frac{s^2\theta^2}{2}\right)}.
\end{equation}
The mean of $f_\theta(\theta)$ is $\langle\theta\rangle=s^{-1}\sqrt{\pi/2}$, giving a mean 
EA of
\begin{equation}
\langle\chi\rangle = \frac{\pi}{4}-\frac{1}{2s}\sqrt{\frac{\pi}{2}}
                   = 45.0^\circ - \frac{35.90^\circ}{s}.
\label{eqn:avg45}
\end{equation}
The mode, or peak, of $f_\theta(\theta)$ is $\theta_p=s^{-1}$. The mode of the EA is then
\begin{equation}
\chi_p = \frac{\pi}{4}-\frac{1}{2s} = 45.0^\circ - \frac{28.65^\circ}{s}.
\label{eqn:mode45}
\end{equation}
The magnitude of the mode is always greater than that of the mean. The mean and mode almost 
never occur at $\chi=\pi/4$, because the instrumental noise in $Q$ and $U$ prevents the 
polarization vector from being precisely aligned with the Stokes $V$ axis of the Poincar\'e 
sphere. 

The standard deviation of $f_\theta(\theta)$ is $\sigma_\theta = s^{-1}\sqrt{2-\pi/2}$. The 
standard deviation of $\chi$ is one-half of this value:
\begin{equation}
\sigma_\chi = \frac{1}{2s}\sqrt{\frac{4-\pi}{2}} = \frac{18.77^\circ}{s}.
\label{eqn:SD45}
\end{equation}


\subsection{Measured EA}
\label{sec:noise}

The measured EA, $\chi_m$, is calculated from the mean linear and circular polarization:
\begin{equation}
\chi_m = \frac{1}{2}\arctan{\left(\frac{\langle V\rangle}{\langle L\rangle}\right)}.
\label{eqn:chi}
\end{equation}
The angular brackets in Equation~\ref{eqn:chi} represent an average of the polarization,
for example, over multiple pulses of a pulsar at a particular pulse longitude or over the 
short duration of each time sample across the pulse of an FRB. For a polarization vector 
with fixed amplitude and orientation, the pdf of the linear polarization is the Rice 
distribution (e.g., Davenport \& Root 1958; Papoulis 1965). The mean linear polarization 
derived from it is (Serkowski 1958)
\begin{equation}
\langle L\rangle 
            = \sigma_n\sqrt{\frac{\pi}{2}}\exp{\left[-\frac{s^2\cos^2(2\chi_o)}{2}\right]}
              {}_1F_1[3/2, 1; s^2\cos^2(2\chi_o)/2].
\label{eqn:Lavg}
\end{equation}
The pdf of the circular polarization is a Gaussian distribution with a mean of 
$\langle V\rangle = \mu_p\sin(2\chi_o)$ and a standard deviation of $\sigma_n$. The measured 
EA is then
\begin{equation}
\chi_m = 
 \frac{1}{2}\arctan{\left\{\sqrt{\frac{2}{\pi}}\exp{\left[\frac{s^2\cos^2(2\chi_o)}{2}\right]}
 \frac{s\sin(2\chi_o)}{{}_1F_1[3/2, 1; s^2\cos^2(2\chi_o)/2]}\right\}}.
\label{eqn:EAfix}
\end{equation}

\begin{figure}
\plotone{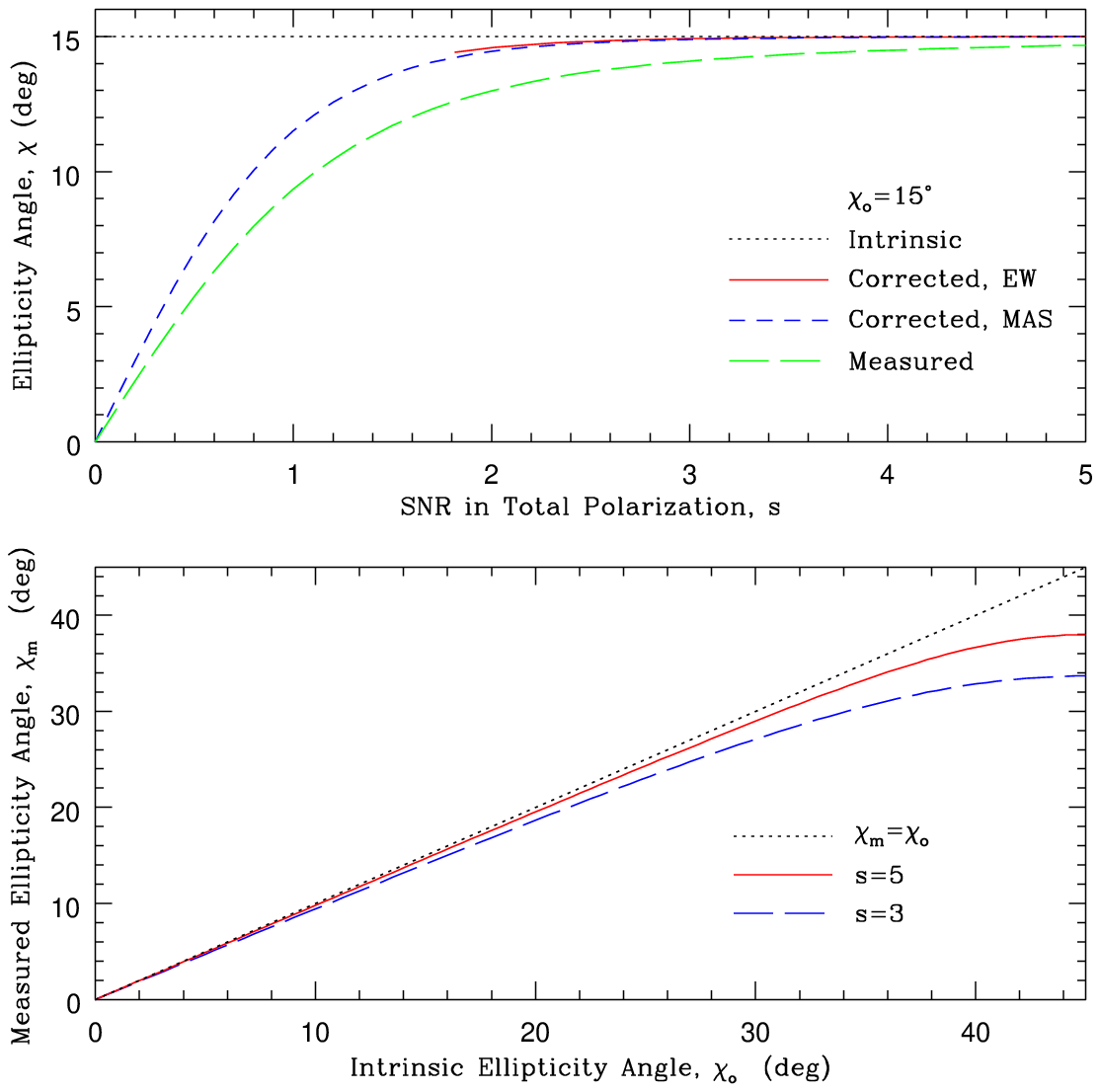}
\caption{Effect of instrumental noise on the measured EA when the amplitude and orientation
of a polarization vector are constant. The dashed green line in the top panel shows the 
dependence of the measured EA upon the SNR in total polarization, $s$, when $\chi_o=15^\circ$
(Equation~\ref{eqn:EAfix}). The solid red line shows the EA as corrected for instrumental 
noise with the EW estimator given by Equation~\ref{eqn:EW}. The dashed blue line shows the 
EA corrected with the MAS estimator given by Equation~\ref{eqn:MAS}. The bottom panel compares 
the measured values of the EA to the intrinsic values for SNRs of $s=3$ (dashed blue line) 
and $s=5$ (solid red line).}
\label{fig:EAfix}
\end{figure}

Examples of $\chi_m$ calculated from Equation~\ref{eqn:EAfix} are shown in Figure~\ref{fig:EAfix}. 
The dashed green line in the top panel of the figure shows the dependence of the measured EA upon 
$s$ when $\chi_o=15^\circ$. The bottom panel compares the measured EA to the intrinsic EA for 
different values of $s$. The measured value of the EA systematically underestimates the intrinsic
value, particularly at low SNR in total polarization and at large values of $\chi_o$. The difference 
between the intrinsic and measured values of the EA is caused by a contribution of the instrumental 
noise to the mean linear polarization. This problem is well known, and a variety of methods have 
been developed to reduce the effect of this instrumental bias (e.g., Simmons \& Stewart 1985;
Everett \& Weisberg 2001; Plaszczynski et al., 2014). The method proposed by Everett \& Weisberg 
(2001), hereafter the EW estimator, is commonly used to compensate the measured linear polarization 
($L_m=\langle L\rangle$) of pulsars for instrumental noise. Their estimate of the true linear 
polarization, $L_t$, from $L_m$ is given by
\begin{equation}
L_t =
\begin{cases}
  \left(L_m^2 - \sigma_n^2\right)^{1/2} & \quad \mathrm{if}\ L_m > 1.57\sigma_n \\
  0 & \quad \mathrm{otherwise}.
\end{cases}
\label{eqn:EW}
\end{equation}
The solid red line in the top panel of Figure~\ref{fig:EAfix} shows the corrected value of 
$\chi_m$ after Equation~\ref{eqn:EW} has been applied to the linear polarization measurement. 
The corrected value slightly underestimates the actual value of $\chi_o$ at low SNR but 
converges  to $\chi_o$ at $s>3$. The corrected value of $\chi_m$ is not shown for $s<1.81$, 
because Equation~\ref{eqn:EW} sets $L_t=0$ for $s\cos(2\chi_o)<1.57$. Plaszczynski et al. 
(2014) proposed a modified asymptotic (MAS) estimator to compensate the measured linear 
polarization for instrumental noise. The MAS estimator calculates $L_t$ from $L_m$ using 
\begin{equation}
L_t = L_m - \frac{\sigma_n^2}{2L_m}
      \left[1-\exp{\left(-\frac{L_m^2}{\sigma_n^2}\right)}\right].
\label{eqn:MAS}
\end{equation}
The dashed blue line in the top panel of Figure~\ref{fig:EAfix} shows the corrected value of 
$\chi_m$ after Equation~\ref{eqn:MAS} has been applied to the linear polarization measurement. 
The corrected value of $\chi_m$ resulting from the application of the MAS estimator is similar 
to that of the EW estimator for $s>2$. It extends to $s=0$, because the MAS estimator attenuates
low values of $L_m$ instead of setting them equal to zero, as with the EW estimator.


\subsection{Standard Deviation of the EA's Probability Density}

\begin{figure}
\plotone{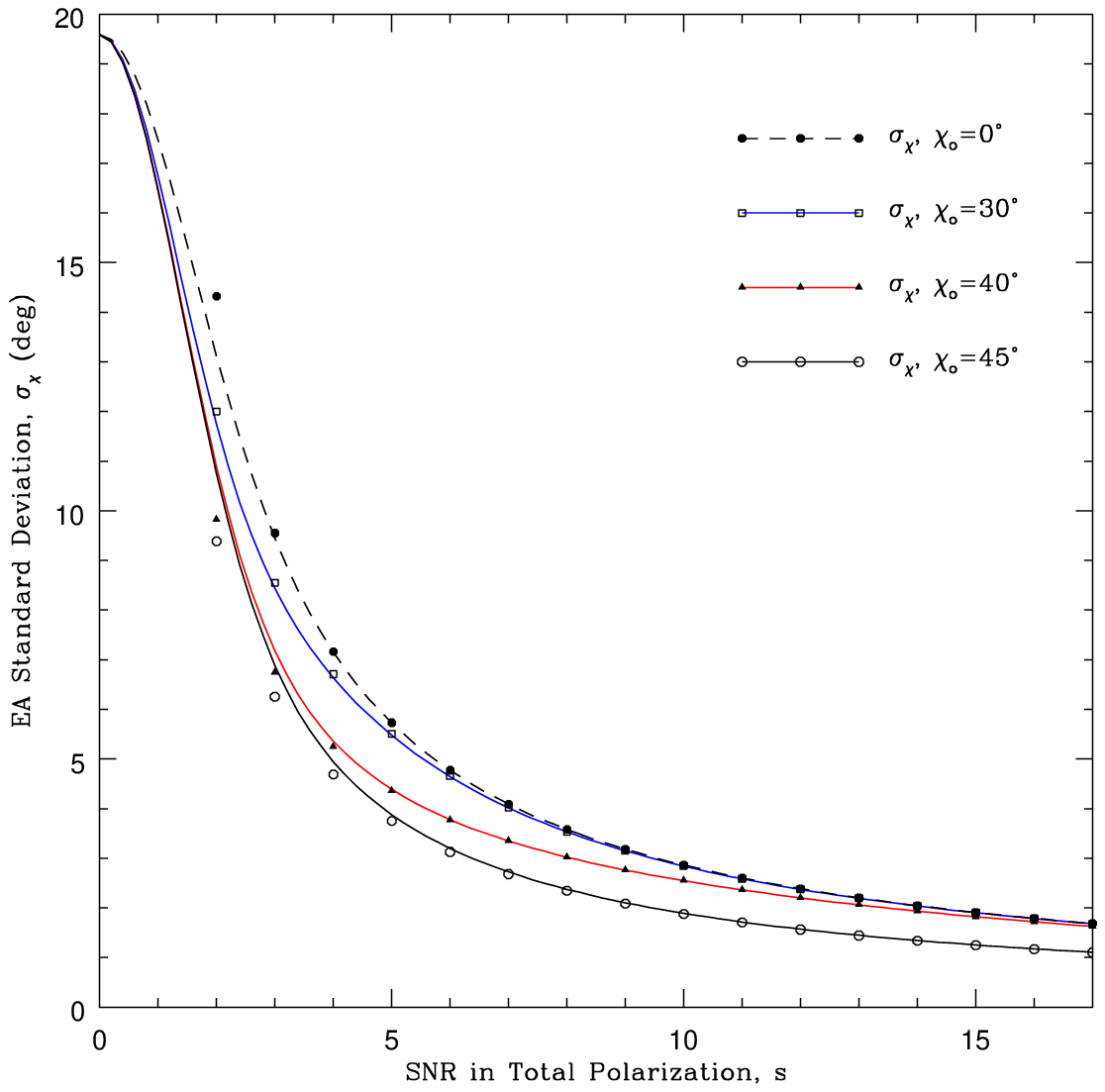}
\caption{Dependence of the standard deviation, $\sigma_\chi$, of $f_\chi(\chi)$ upon the SNR 
in total polarization, $s$. The black, blue, and red lines show $\sigma_\chi$ calculated 
from the first and second moments of $f_\chi(\chi)$ for different values of $\chi_o$ as 
annotated in the figure. The symbols attached to each line are approximations to 
$\sigma_\chi$ as discussed in the text.}
\label{fig:chiSD}
\end{figure}

The standard deviation of the EA, $\sigma_\chi$, can be calculated numerically from the EA 
pdfs derived in Section~\ref{sec:chiPDF}. Unlike the standard deviation of the PA, which is 
independent of $\psi_o$, $\sigma_\chi$ is generally dependent upon $\chi_o$. Examples of 
$\sigma_\chi$ as functions of $s$ for different values of $\chi_o$ are illustrated in 
Figure~\ref{fig:chiSD}. The figure shows all values of $\sigma_\chi$ are confined to an envelope 
bounded by the curves representing $\sigma_\chi$ for $\chi_o=0^\circ$ and $\chi_o=45^\circ$. The 
curves representing $\sigma_\chi$ for $\chi_o=30^\circ$ and $\chi_o=40^\circ$ initially trace the 
lower boundary of the envelope at low SNR but migrate to the upper boundary as the SNR increases. 
The curves for a small value of $\chi_o$ merge with the upper boundary at a smaller value of $s$
than the curves for a large value of $\chi_o$. This migration between curves is a manifestation of 
$f_\chi(\chi)$ evolving from an asymmetric distribution to a symmetric one as the SNR increases.
Regardless of the value of $\chi_o$, $\sigma_\chi$ converges to the standard deviation of the 
$\cos(2\chi)$ distribution at $s=0$:
\begin{equation}
\sigma_\chi = \frac{1}{4}\sqrt{\pi^2-8}=19.59^\circ.
\end{equation}

Approximations to $\sigma_\chi$ can be determined from the definition of $\chi$ 
(Equation~\ref{eqn:chi}) using the standard propagation of errors (e.g., Naghizadeh-Khouei 
\& Clarke 1993; Cao et al. 2025):
\begin{equation}
\sigma_\chi = \frac{(L^2\sigma_v^2 + V^2\sigma_l^2)^{1/2}}{2P^2}
            = \frac{[\cos^2(2\chi_o)\sigma_v^2 + \sin^2(2\chi_o)\sigma_l^2]^{1/2}}{2P}.
\label{eqn:SDapprox}
\end{equation}
%
%
%
Here, $P=\sqrt{L^2+V^2}$ is the total polarization, $\sigma_v^2=\sigma_n^2$ is the variance 
of the circular polarization, and $\sigma_l^2$ is the variance of the linear polarization 
calculated from the Rice distribution:
\begin{equation}
\sigma_l^2 = \sigma_n^2\Biggl\{2 + s^2\cos^2(2\chi_o)
           - \frac{\pi}{2}\Biggl\{\exp{\left[-\frac{s^2\cos^2(2\chi_o)}{2}\right]}
                 {}_1F_1[3/2,1;s^2\cos^2(2\chi_o)/2]\Biggr\}^2\Biggr\}.
\end{equation}
%
%
%
Equation~\ref{eqn:SDapprox} shows that $\sigma_\chi$ is a function of the intrinsic EA. For 
linearly polarized radiation ($\chi_o=0$), the approximation to $\sigma_\chi$ calculated
from Equation~\ref{eqn:SDapprox} is the same result given by Equation~\ref{eqn:chiSD0}. 
The approximation is shown by the filled circles in Figure~\ref{fig:chiSD}. It reproduces
the actual standard deviation for SNRs of $s>3$. The approximation is also applicable to 
all combinations of $s$ and $\chi_o$ satisfying the condition $s\cos(2\chi_o)\gg\sqrt{2}$. 
This condition indicates where the curves for $\sigma_\chi$ coincide with the upper boundary 
of the envelope. It may be interpreted as the combination of $s$ and $\chi_o$ that defines 
where $f_\chi(\chi)$ transitions from an asymmetric pdf to a symmetric one. The actual merger 
point in the figure, and thus the condition for when the approximation to a symmetric pdf is 
applicable, is roughly $s > 3/\cos(2\chi_o)$.

For circularly polarized radiation ($\chi_o=\pi/4$), $\sigma_l=\sigma_n\sqrt{2-\pi/2}$,
and the approximation calculated from Equation~\ref{eqn:SDapprox} is the same result 
as given by Equation~\ref{eqn:SD45}. The approximation is shown by the open circles in 
Figure~\ref{fig:chiSD}. It reproduces the actual standard deviation for SNRs of $s>4$.

Equation~\ref{eqn:SDapprox} was also used to calculate the approximations to $\sigma_\chi$ for 
$\chi_o=30^\circ$ and $40^\circ$. The approximations are shown by the open squares and filled
triangles, respectively, in Figure~\ref{fig:chiSD}. These approximations also reproduce the 
actual values of $\sigma_\chi$ for SNRs of $s>4$.

While the standard deviation can be used as an estimate of the error in an EA measurement,
it should not be misconstrued as a confidence limit, because the standard deviation is 
equal to a $68\%$ confidence limit only when the parent pdf is Gaussian. Additionally, the 
EA pdf is generally asymmetric, which means its upper and lower confidence limits are not 
equal. A method for calculating the EA confidence limits is proposed in the Appendix.


\section{EA from Polarization Fluctuations}
\label{sec:opm}

\subsection{OPM Statistical Model}

OPMs frequently occur in the radio emission of pulsars (Manchester et al., 1975; Cordes et al., 
1978; Backer \& Rankin 1980; Stinebring et al., 1984). The modes appear as abrupt discontinuities 
in PA of $\Delta\psi\simeq\pi/2$ in both average profiles and single pulses. The emission tends 
to be depolarized where the PA discontinuities occur. OPMs can also appear as bimodal histograms 
of the PA at some longitudes of a pulsar's pulse. Cordes et al. (1978) showed that the mode PAs 
in PSR B2020+28 are correlated with the handedness of circular polarization, indicating the 
modes are generally elliptically polarized and their polarization vectors are directed toward 
antipodal points on the Poincar\'e sphere. The switching between modes is likely related to some 
type of stochastic process (Cordes 1981). 

McKinnon \& Stinebring (1998, 2000) developed a statistical model for the polarization of 
the emission that is based on these observations. The model simulates the pulse-to-pulse 
fluctuations in polarization at a particular pulse longitude, as opposed to variations in 
the average polarization across a pulse profile. The model assumes the OPMs are incoherent, 
superposed, and completely polarized and treats the mode intensities as random variables to 
account for the random switching between orthogonally polarized states. With the variables 
$X_A$ and $X_B$ representing the mode intensities, the Stokes parameters produced by the model 
are (e.g., McKinnon 2003)
\begin{equation}
Q = \cos(2\psi_o)\cos(2\chi_o)(X_A - X_B) + X_{N,Q},
\label{eqn:Q}
\end{equation}
\begin{equation}
U = \sin(2\psi_o)\cos(2\chi_o)(X_A - X_B) + X_{N,U},
\end{equation}
\begin{equation}
V = \sin(2\chi_o)(X_A - X_B) + X_{N,V}.
\label{eqn:V}
\end{equation}
Here, the random variables $X_{N,Q}$, $X_{N,U}$, and $X_{N,V}$ represent the independent 
instrumental noise in each of the measured Stokes parameters $Q$, $U$, and $V$. The magnitude 
of these random variables is assumed to be equal, with a value of $\sigma_n$. The equations 
are written assuming mode A is the stronger of the two modes. Therefore, the PA of mode A is 
$\psi_o$, and its EA is $\chi_o$. The PA of mode B is $\psi_o\pm\pi/2$ and its EA is $-\chi_o$. 
The fluctuations in mode intensities cause the amplitude of the resultant polarization vector 
to fluctuate. Since the OPMs are assumed to be superposed and incoherent, the average SNR in 
total polarization is determined by the difference in the mode mean intensities. Therefore, 
while the SNR of a single polarization sample can be very large, the SNR on average will be 
small when the mode mean intensities are comparable. 

\subsection{The Q-V Correlation Coefficient} 

Equations~\ref{eqn:Q}-\ref{eqn:V} show that the Stokes parameters are covariant, because 
they all vary as $X_A-X_B$. The covariance must be accounted for in constructing the joint 
probability density of the polarization vector's amplitude and orientation angles. When the 
intrinsic PA is $\psi_o=0$, the polarized signal is absent from the Stokes parameter $U$, 
and the correlation occurs between the Stokes parameters $Q$ and $V$. Assuming the standard 
deviations of the mode intensities are equal, $\sigma_A=\sigma_B=\sigma$, the standard 
deviations of $Q$ and $V$ are
\begin{equation}
\sigma_q = \sigma_n[1+\rho^2\cos^2(2\chi_o)]^{1/2},
\end{equation}
\begin{equation}
\sigma_v = \sigma_n[1+\rho^2\sin^2(2\chi_o)]^{1/2},
\end{equation}
where $\rho=\sigma\sqrt{2}/\sigma_n$ represents the intrinsic fluctuations in the amplitude of 
the polarization vector relative to the instrumental noise. The covariance of $Q$ and $V$ can 
be derived from their definitions given by Equations~\ref{eqn:Q} and~\ref{eqn:V}:
\begin{equation}
{\rm Cov}(Q,V) = \cos(2\chi_o)\sin(2\chi_0)\rho^2\sigma_n^2.
\end{equation}

\begin{figure}
\plotone{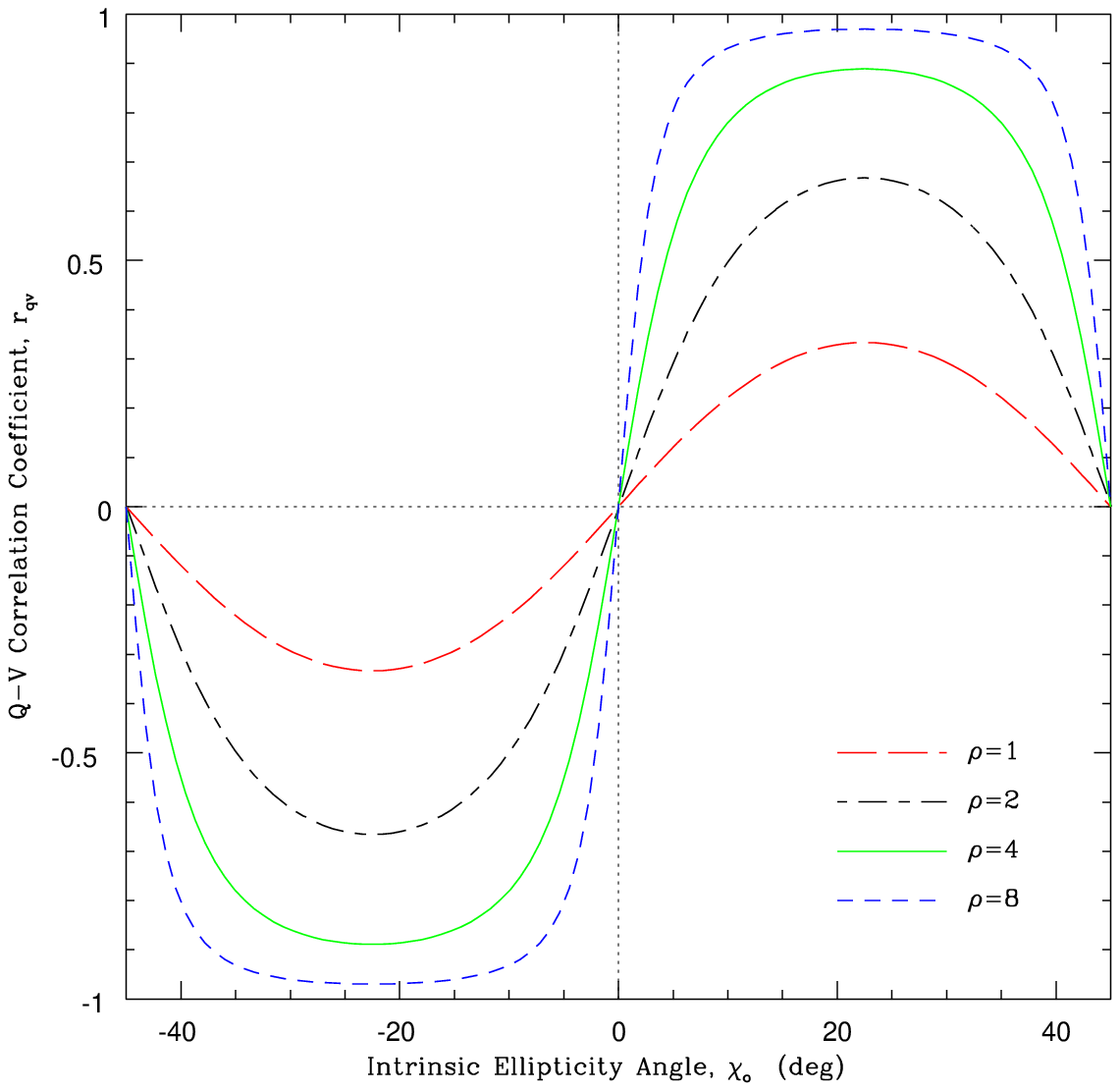}
\caption{Dependence of the $Q$-$V$ correlation coefficient, $r_{qv}$, upon the intrinsic
EA, $\chi_o$, and polarization fluctuation ratio, $\rho$.}
\label{fig:rQV}
\end{figure}

The resulting $Q$-$V$ correlation coefficient is
\begin{equation}
r_{qv} = \frac{{\rm Cov}(Q,V)}{\sigma_q\sigma_v}
           = \frac{\rho^2\sin(4\chi_o)}{[4(1+\rho^2) + \rho^4\sin^2(4\chi_o)]^{1/2}}.
\label{eqn:rQV}
\end{equation}
The variation of $r_{qv}$ with $\chi_o$ for different values of $\rho$ is shown in 
Figure~\ref{fig:rQV}. The coefficient is antisymmetric about $\chi_o=0$ with $Q$ and 
$V$ being correlated for $\chi_o>0$ and anticorrelated for $\chi_o<0$. The magnitude 
of $r_{qv}$ increases with increasing $\rho$ and is maximum at $\chi_o=\pm\pi/8$, 
where its value is $r_{qv}=\pm\rho^2/(2+\rho^2)$. The Stokes parameters $Q$ and 
$V$ are independent ($r_{qv}=0$) when $\rho=0$, $\chi_o=\pm\pi/4$, or $\chi_o=0$. 
In the case of $\rho=0$, the amplitude of the polarization vector is constant, and the 
only fluctuations in the recorded signal are due to instrumental noise. This case was 
evaluated in Section~\ref{sec:fix}. For the case of $\chi_o=\pm\pi/4$, the emission 
is circularly polarized, and no signal occurs within $Q$. This case was examined in 
McKinnon (2006), albeit in terms of the polarization vector's colatitude instead of
its EA. The pdf of the colatitude is given by Equation 6 of McKinnon (2006), and 
examples of it are shown in Figure 1 of that paper. For the case of $\chi_o=0$, the 
emission is linearly polarized, and $V$ is comprised solely of instrumental noise. 

Equation~\ref{eqn:rQV} is general, in the sense that no assumption has been made regarding 
the statistical character of the mode intensity fluctuations in its derivation. The mode 
intensities are assumed to be Gaussian random variables in the analysis that follows.


\subsection{General Joint Probability Density}

Following the procedure outlined in Section~\ref{sec:fixjoint}, the general expression for the 
joint probability density of $\psi$ and $\chi$ can be derived from the $Q$-$V$ correlation 
coefficient and the definitions of the Stokes parameters given by 
Equations~\ref{eqn:Q}-\ref{eqn:V}. The joint density is
\begin{eqnarray}
f(\psi,\chi) & = & \frac{\cos(2\chi)}{\pi}\exp{\left(-\frac{s^2}{2\sigma_\rho^2}\right)}
                   \frac{\sigma_\rho^2}{[\sigma_\rho^2 - \rho^2g^2(\psi,\chi)]^{3/2}}
                   \Biggl\{h(\psi,\chi)\sqrt{\frac{2}{\pi}} 
                 + [1+h^2(\psi,\chi)]\exp{\left[\frac{h^2(\psi,\chi)}{2}\right]} \nonumber \\
        & \times & \Biggl\{1+{\rm erf}\left[\frac{h(\psi,\chi)}{\sqrt{2}}\right]\Biggr\}\Biggr\},
\label{eqn:genjoint}
\end{eqnarray}
%
%
%
where $\sigma_\rho = \sqrt{1 + \rho^2}$ and $h(\psi,\chi)$ is 
\begin{equation}
h(\psi,\chi)=\frac{sg(\psi,\chi)}{\sigma_\rho [\sigma_\rho^2-\rho^2g^2(\psi,\chi)]^{1/2}}.
\end{equation}
When $\rho=0$, the amplitude of the polarization vector is constant, $h(\psi,\chi)=sg(\psi,\chi)$,
and Equation~\ref{eqn:genjoint} reduces to Equation~\ref{eqn:fixjoint}.

\begin{figure}
\plotone{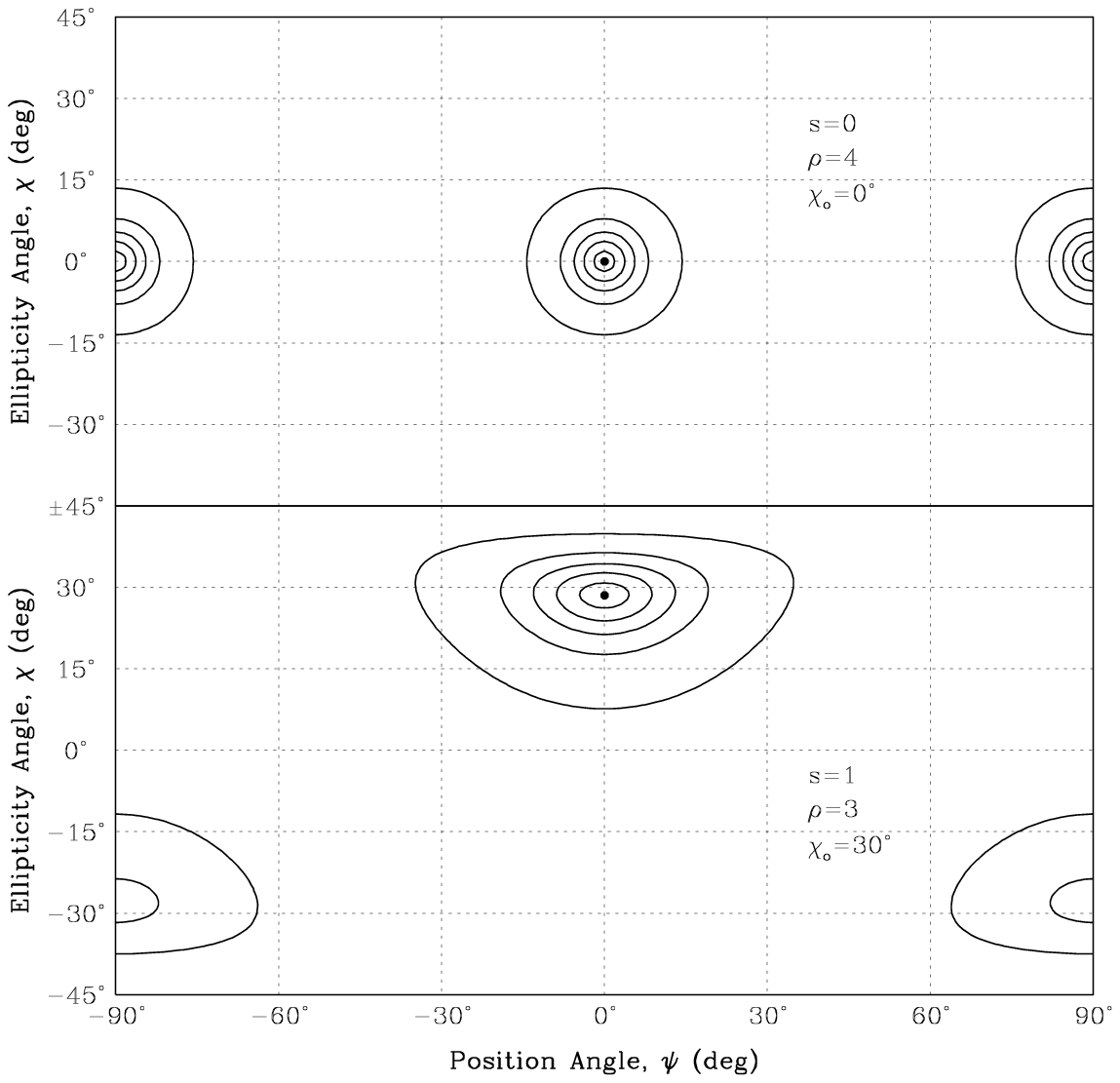}
\caption{Examples of the joint probability density, $f(\psi,\chi)$, of the PA and EA when
the amplitude of the polarization vector fluctuates due to the superposition of incoherent
OPMs (Equation~\ref{eqn:genjoint}). The top panel shows $f(\psi,\chi)$ when $s=0$, $\rho=4$, 
and $\chi_o=0^\circ$. The dot marks the peak of $f(\psi,\chi)$ at $\psi=\chi=0^\circ$. The 
bottom panel shows $f(\psi,\chi)$ when $s=1$, $\rho=3$, and $\chi_o=30^\circ$. The dot marks 
the peak of $f(\psi,\chi)$ at $\psi=0^\circ$ and $\chi=28.5^\circ$. The contour levels in 
both panels are drawn at 0.1, 0.3, 0.5, 0.7, and 0.9 of the peak value.} 
\label{fig:genjoint}
\end{figure}

Two examples of Equation~\ref{eqn:genjoint} are shown in Figure~\ref{fig:genjoint}. In both 
cases, the joint density is bimodal, due to the fluctuations in the intensities of the OPMs. 
The top panel of the figure shows $f(\psi,\chi)$ when $s=0$, $\rho=4$, and $\chi_o=0^\circ$. 
In this example, each component of the joint density is nearly circularly symmetric. The 
bottom panel of the figure shows $f(\psi,\chi)$ when $s=1$, $\rho=3$, and $\chi_o=30^\circ$. 
The component at $\psi=0^\circ$ and $\chi\simeq 30^\circ$ is comprised of data samples with 
$X_A > X_B$. Similarly, the component at $\psi=\pm 90^\circ$ and $\chi\simeq -30^\circ$ is 
comprised of data samples with $X_A < X_B$. The component at $\psi=0^\circ$ is the more 
prominent of the two, because the intensity of mode A is greater than that of mode B on 
average, due to $s=1$. Each component is symmetric in PA and asymmetric in EA. The shapes 
of the two components become mirror images of one another about $\chi=0^\circ$ when $s=0$ 
and are identical and completely symmetric when $s=0$, $\chi_o=0^\circ$, and $\rho\gg 1$. 
The joint density becomes unimodal when the mean amplitude of the polarization vector 
exceeds the polarization fluctuations (i.e., when $s>\sqrt{1+\rho^2}$).
 

\subsection{General Probability Density of the EA}
\label{sec:eagen}

\begin{figure}
\plotone{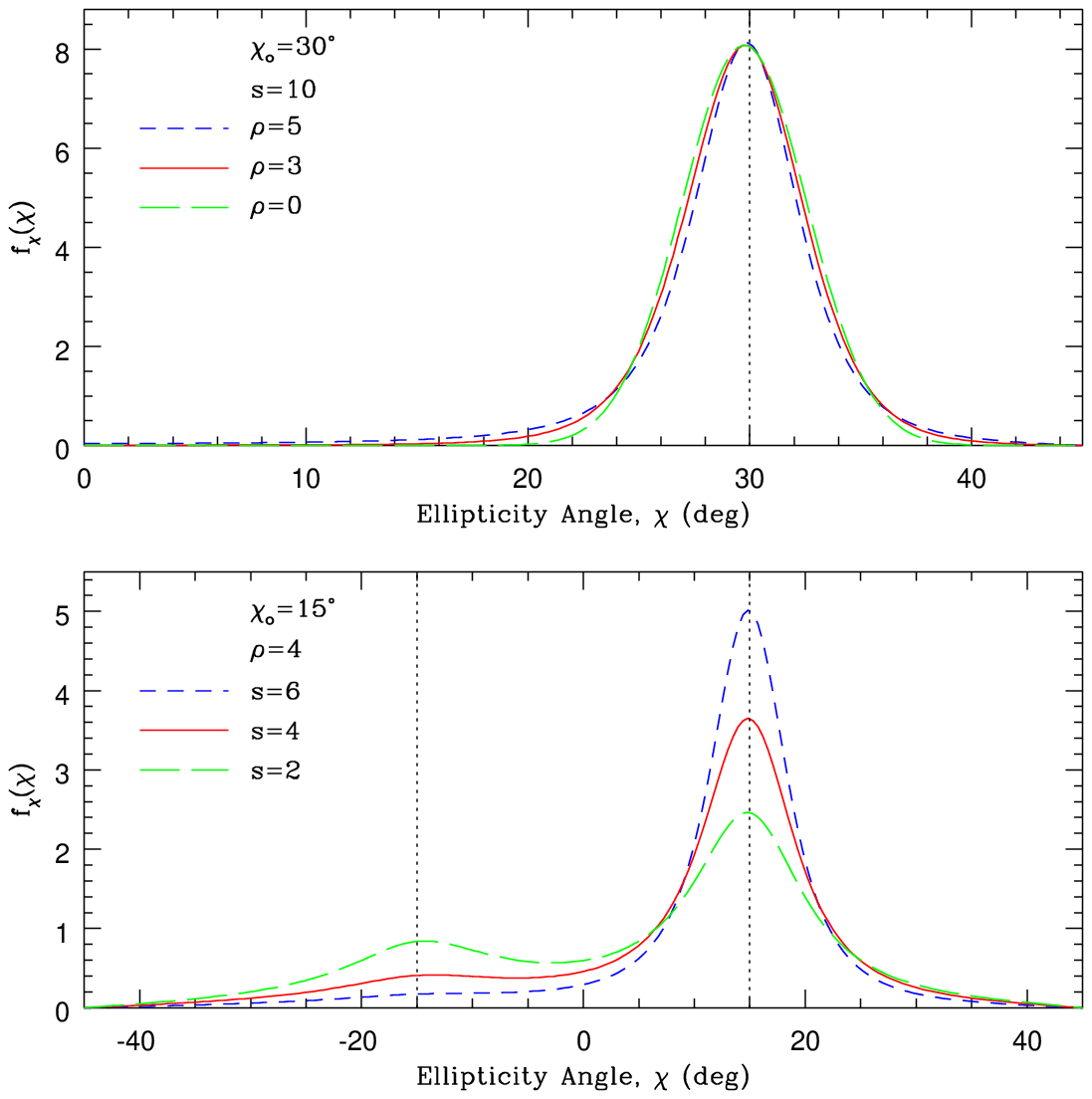}
\caption{Examples of the general pdf of the EA for incoherent OPMs. The top panel shows 
examples of the pdf for different values of the polarization fluctuation ratio, $\rho$, 
when the intrinsic EA and SNR are held constant at $\chi_o=30^\circ$ and $s=10$, 
respectively. The bottom panel shows example pdfs for different values of SNR when the 
intrinsic EA and the polarization fluctuation ratio are held constant at $\chi_o=15^\circ$ 
and $\rho=4$. The dotted vertical lines denote the intrinsic EAs of the OPMs, $\pm\chi_o$.}
\label{fig:genchi}
\end{figure}

As in Section~\ref{sec:chiPDF}, the general pdf of $\chi$ can be found by numerically 
integrating the joint probability density given by Equation~\ref{eqn:genjoint} over $\psi$. 
Examples of this general pdf are shown in Figure~\ref{fig:genchi}. The bottom panel of the 
figure shows the pdf for different values of $s$ when the intrinsic EA of the dominant 
polarization mode and the polarization fluctuation ratio are held constant at $\chi_o=15^\circ$ 
and $\rho=4$, respectively. The pdf is bimodal when the polarization fluctuations exceed the 
mean polarization ($s<\sqrt{1+\rho^2}$) and becomes unimodal as the mean polarization exceeds 
the fluctuations ($s>\sqrt{1+\rho^2}$). The top panel of the figure shows the pdf when the 
mean polarization is much larger than the fluctuations. These pdfs have different values of 
$\rho$, but the same values of the intrinsic EA ($\chi_o=30^\circ$) and polarization SNR 
($s=10$). The pdfs are unimodal and almost indistinguishable from one another. (Note that 
the range of the abscissa in the top panel is half that of the bottom panel.) The pdf with
$\rho=0$ is nearly symmetric, while the other pdfs have tails that extend toward $\chi=0$.
The pdf becomes Gaussian as the mean polarization becomes even larger with respect to the 
polarization fluctuations. The height of the Gaussian pdf increases and its width decreases 
with increasing SNR. In this instance, the emission is dominated by one polarization mode.
 

\subsection{Probability Density of the EA for OPMs with Equal Mean Intensities}

\begin{figure}
\plotone{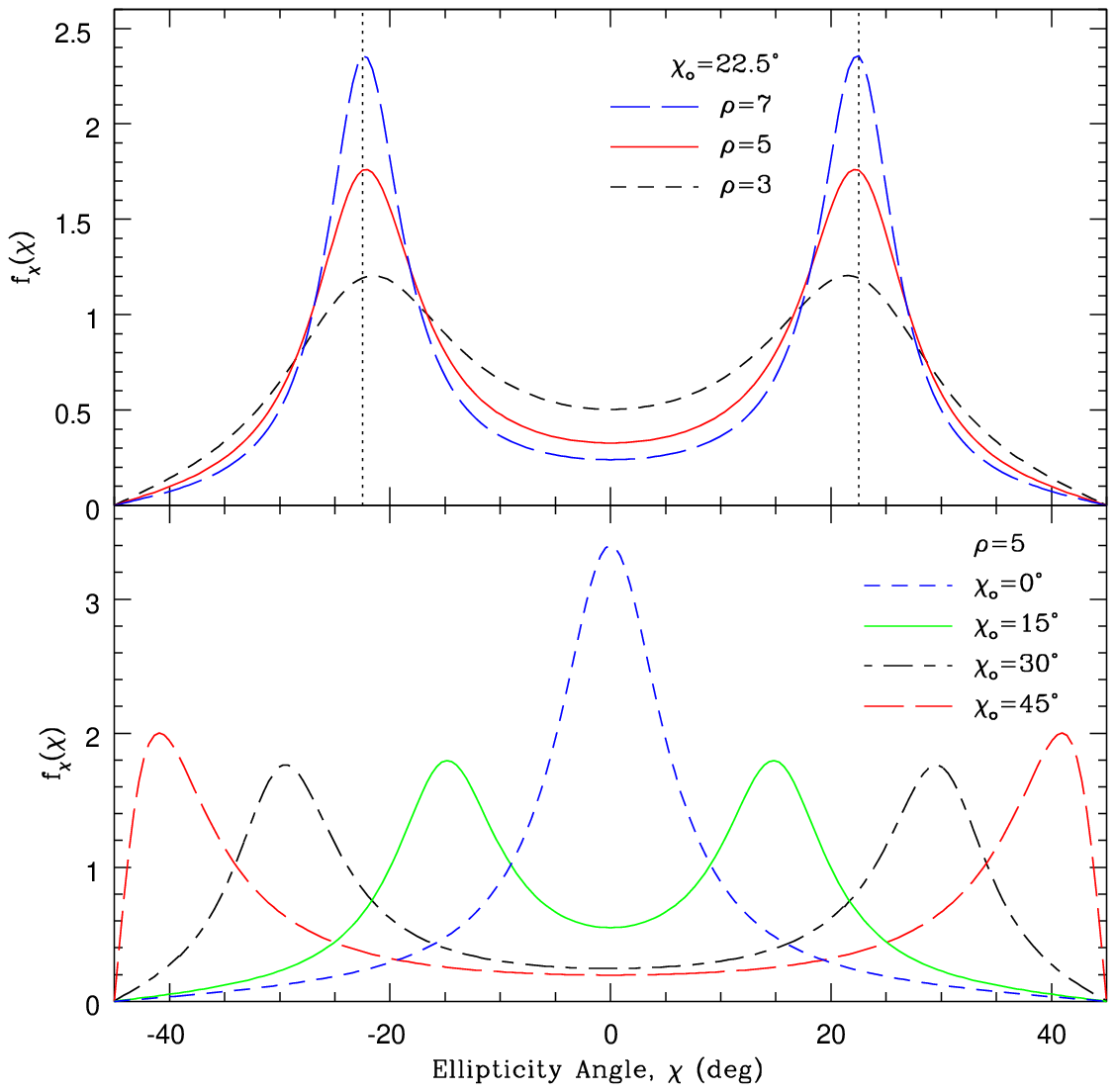}
\caption{Example pdfs of the EA for incoherent OPMs with the same mean intensity ($s=0$). 
The top panel shows examples of the pdf for different values of the polarization fluctuation 
ratio, $\rho$, when the intrinsic EA is held constant at $\chi_o=22.5^\circ$. The dotted 
vertical lines denote the intrinsic EAs of the OPMs, $\pm\chi_o$. The bottom panel shows 
example pdfs for different values of $\chi_o$ when the polarization fluctuation ratio 
remains fixed at $\rho=5$.}
\label{fig:chidist}
\end{figure}

When the mean intensities of the modes are equal, $\langle X_A\rangle=\langle X_B\rangle$,
the mean polarization is equal to zero ($s=0$), and the joint probability density of 
$\psi$ and $\chi$ is
\begin{equation}
f(\psi,\chi) = \frac{\cos(2\chi)}{\pi}\frac{1+\rho^2}{[(1+\rho^2)-\rho^2g^2(\psi,\chi)]^{3/2}}.
\label{eqn:opmjoint}
\end{equation}
Again, the pdf of $\chi$ can be found by integrating the joint probability density given by 
Equation~\ref{eqn:opmjoint} over $\psi$. The pdf can be determined analytically when the 
intrinsic EA is $\chi_o=0$ or $\chi_o=\pi/4$ and can be calculated numerically for other 
values of $\chi_o$. The top panel of Figure~\ref{fig:chidist} shows the pdf when 
$\chi_o=\pi/8$ for different values of $\rho$. The pdf is generally bimodal. The separation 
of the pdf's component peaks is generally less than $2\chi_o$, but approaches $2\chi_o$ as 
$\rho$ increases. The width of each component also decreases as $\rho$ increases. The bottom 
panel of the figure shows examples of the pdf for different values of $\chi_o$ with the 
polarization fluctuation ratio fixed at $\rho=5$. The two components of $f_\chi(\chi)$ are 
generally asymmetric about their peaks, and their separation tends to be less than $2\chi_o$, 
particularly at large values of $\chi_o$. 

For the specific case of $\chi_o=0$, the pdf of $\chi$ is
\begin{eqnarray}
f_\chi(\chi) & = & \cos(2\chi)(1+\rho^2)
                   \left[\frac{2}{2+\rho^2(1+\sin^2(2\chi))}\right]^{3/2} \nonumber \\
& \times & {}_2F_1\left\{3/4, 5/4; 1; \rho^4\cos^4(2\chi)/[2+\rho^2(1+\sin^2(2\chi)]^2\right\}.
\label{eqn:chi0}
\end{eqnarray}
The function ${}_2F_1(a,b;c; z)$ in Equation~\ref{eqn:chi0} is the Gauss hypergeometric 
function, with numerical parameters $a$, $b$, and $c$ and argument $z$. This pdf is unimodal,
with a peak at $\chi=0$. The full width of the EA pdf at half its maximum value (FWHM) 
approaches $w_{50}=\arcsin(1/\rho)$ as $\rho$ becomes large. An example of this pdf is 
shown by the dashed blue line in the bottom panel of Figure~\ref{fig:chidist}. The pdf 
of the PA that accompanies $f_\chi(\chi)$ in this case is (Equation 17 of McKinnon \& 
Stinebring 1998)
\begin{equation}
f_\psi(\psi)=\frac{1}{\pi}\frac{(1+\rho^2)^{1/2}}{[1+\rho^2\sin^2(2\psi)]}.
\end{equation}
The PA pdf is also bimodal. The two components of this pdf are separated by 
$\Delta\psi=\pi/2$, and the FWHM of each component is also $w_{50}=\arcsin(1/\rho)$ for 
$\rho\ge 1$. Therefore, each component in the joint density of the polarization angles is 
circularly symmetric, as shown in the top panel of Figure~\ref{fig:genjoint}.

\begin{figure}
\plotone{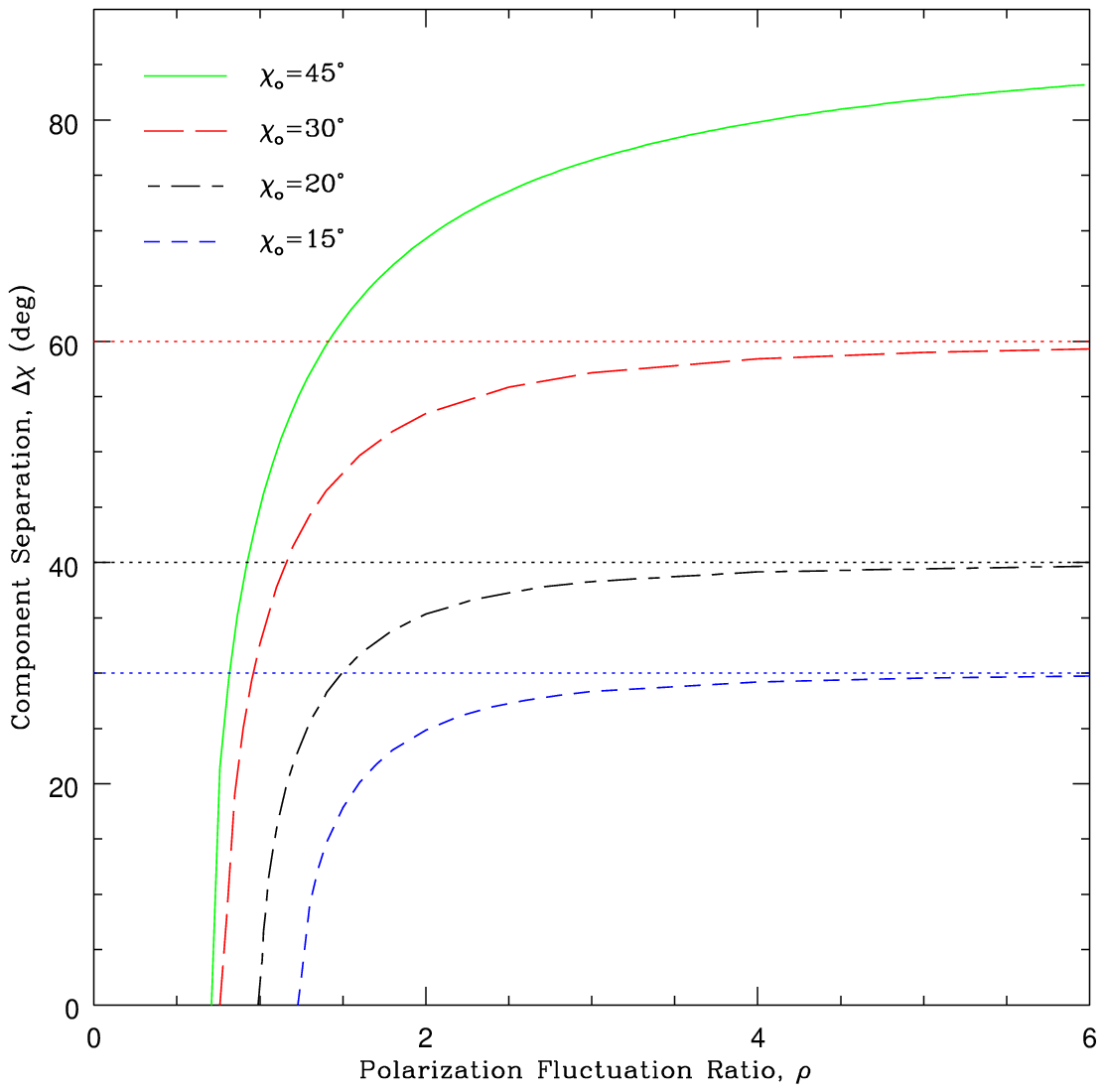}
\caption{Dependence of the component separation in the EA pdf upon the polarization 
fluctuation ratio, $\rho$, for different values of the intrinsic EA, $\chi_o$. The dotted 
horizontal lines denote the actual difference in mode EAs, $2\chi_o$. In each case, the 
component separation asymptotically approaches $2\chi_o$ as $\rho$ increases.}
\label{fig:sep}
\end{figure}

When $\chi_o=\pi/4$, $f_\chi(\chi)$ is given by 
\begin{equation}
f_\chi(\chi)=\cos(2\chi)\frac{1+\rho^2}{[1+\rho^2\cos^2(2\chi)]^{3/2}}.
\end{equation}
An example of this pdf is shown by the dashed red line in the bottom panel of 
Figure~\ref{fig:chidist}. The two components of the pdf are asymmetric about their peaks.
The angular separation of the peaks is
\begin{equation}
\Delta\chi = \arccos\left(\frac{1}{\rho\sqrt{2}}\right).
\label{eqn:sep45}
\end{equation}
The dependence of $\Delta\chi$ upon $\rho$ given by Equation~\ref{eqn:sep45} is shown by 
the solid green line in Figure~\ref{fig:sep}. The component separation gradually approaches 
$\Delta\chi=\pi/2$ as $\rho$ increases. The components merge into a unimodal pdf centered 
on $\chi=0$ when the polarization fluctuations decrease to a critical value of 
$\rho_c=1/\sqrt{2}$. The pdf remains unimodal for values of $\rho$ that are less than this 
critical value. The dependence of the component separation upon $\rho$ for other values of 
$\chi_o$ is also shown in Figure~\ref{fig:sep}. The behavior of $\Delta\chi$ is similar in 
all cases. The component separation approaches $2\chi_o$ as $\rho$ increases and abruptly 
decreases to zero at a critical value of $\rho$ where $f_\chi(\chi)$ becomes unimodal. 
The critical value of $\rho$ increases with decreasing $\chi_o$. From an observational 
perspective, the figure shows that the polarization fluctuations relative to the 
instrumental noise must be sufficiently large for the components of the bimodal pdf to be 
resolved.

Equations 27 and 28 of McKinnon (2003) attempt to define a general expression for the joint 
probability density of the polarization vector's azimuth ($\phi=2\psi$) and colatitude, 
$\theta$, when $s=0$. However, the equations are incorrect, because the covariance of the 
Stokes parameters was neglected in their derivation. The correct general expression for the 
joint density of the orientation angles when $s=0$ is given by Equation~\ref{eqn:opmjoint} 
of this paper.


\section{Comparison with Observations}
\label{sec:obs}

Both the joint probability density of the polarization angles and the EA pdf derived 
in Section~\ref{sec:opm} can be compared to the results of single-pulse polarization 
observations of pulsars to determine if they are reasonable representations of what is 
actually observed.

\subsection{Observed Probability Density of the EA}

\begin{figure}
\plotone{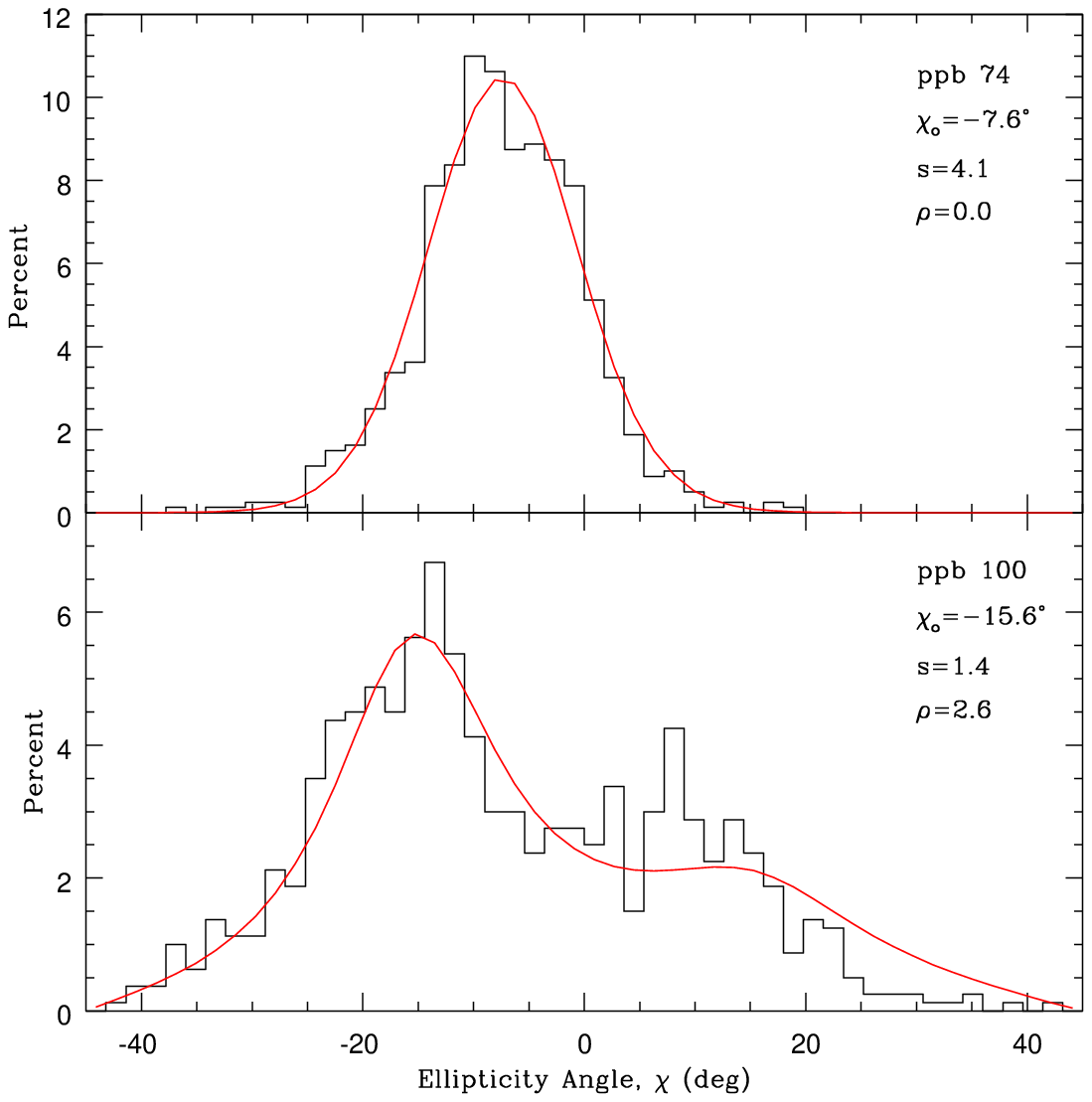}
\caption{Comparison of theoretical EA pdfs with histograms of the EA observed in 
PSR B2020+28 at 1404 MHz. The top panel shows the EA histogram recorded near the center 
of the pulse at pulse phase bin (ppb) 74 in Figure 2a of McKinnon (2004). The continuous 
red line represents the result of a chi-squared fit of the histogram to the theoretical 
pdf derived numerically from Equation~\ref{eqn:genjoint}. The parameters of the fit are 
annotated in the panel. The bottom panel compares the EA histogram recorded at the 
trailing peak of the pulse (ppb 100) with its best-fit pdf.}
\label{fig:eahist}
\end{figure}

Figure~\ref{fig:eahist} compares the theoretical EA pdfs derived numerically from 
Equation~\ref{eqn:genjoint} with EA histograms observed at two locations within the pulse 
of PSR B2020+28. The polarization observations were made by Stinebring et al. (1984) with 
the Arecibo radio telescope at a frequency of 1404 MHz. Each histogram in the figure is 
comprised of 800 data samples. The continuous red line in each panel of the figure 
represents the result of a chi-squared fit of the observed histogram to the theoretical 
pdf. The parameters determined by the fits are annotated in the panels. The red lines are 
credible reproductions of the observed histograms. The EA histogram in the top panel of 
the figure was constructed from data samples recorded near the center of the pulse at 
phase bin 74 in Figure 2a of McKinnon (2004). The histogram is unimodal, broad, and 
centered on an EA of $\chi=-7.6^\circ$. The histogram's width can be attributed to the 
moderate SNR of $s=4.1$ found in the fit to the data. The value of $\rho=0.0$ determined 
by the fit is consistent with McKinnon's (2004) interpretation that the amplitude of the 
polarization vector in this region of the pulse is constant. The histogram in the bottom 
panel of the figure was constructed from data samples recorded at the peak of the trailing 
component in the pulse (at phase bin 100 in Figure 2a of McKinnon 2004), where the 
magnitude of the average circular polarization attains its highest value within the pulse. 
The histogram is very broad and bimodal, with one polarization mode being stronger than 
the other. The fit to the histogram suggests the intrinsic EA of the dominant mode is 
$\chi_o=-15.6^\circ$ and indicates the polarization fluctuations ($\rho=2.6$) exceed 
the mean polarization ($s=1.4$).

\subsection{Observed Joint Probability Density of the Polarization Angles}

Dyks et al. (2021) conducted single-pulse polarization observations of PSR B1451-68 
with the Parkes radio telescope at a frequency of 1369 MHz. Their Figure 1 shows OPMs 
occurring across the entirety of the pulsar's pulse. The mode EAs also vary systematically 
between $\chi\simeq -\pi/4$ and $\chi\simeq\pi/4$, tracing a bow tie shape across the pulse. 
The figure also reveals polarization properties that are consistent with the OPM statistical 
model summarized in Section~\ref{sec:opm}. Specifically, the instantaneous polarization 
fraction of the emission can be high, while the polarization fraction on average is low. 
Their Figure 2 shows color-coded Hammer projections of the PAs and EAs recorded at specific 
pulse longitudes on the Poincar\'e sphere. The projections are generally bimodal, suggesting 
that the polarization fluctuations exceed the mean polarization almost everywhere within the 
pulse.

\begin{figure}
\plotone{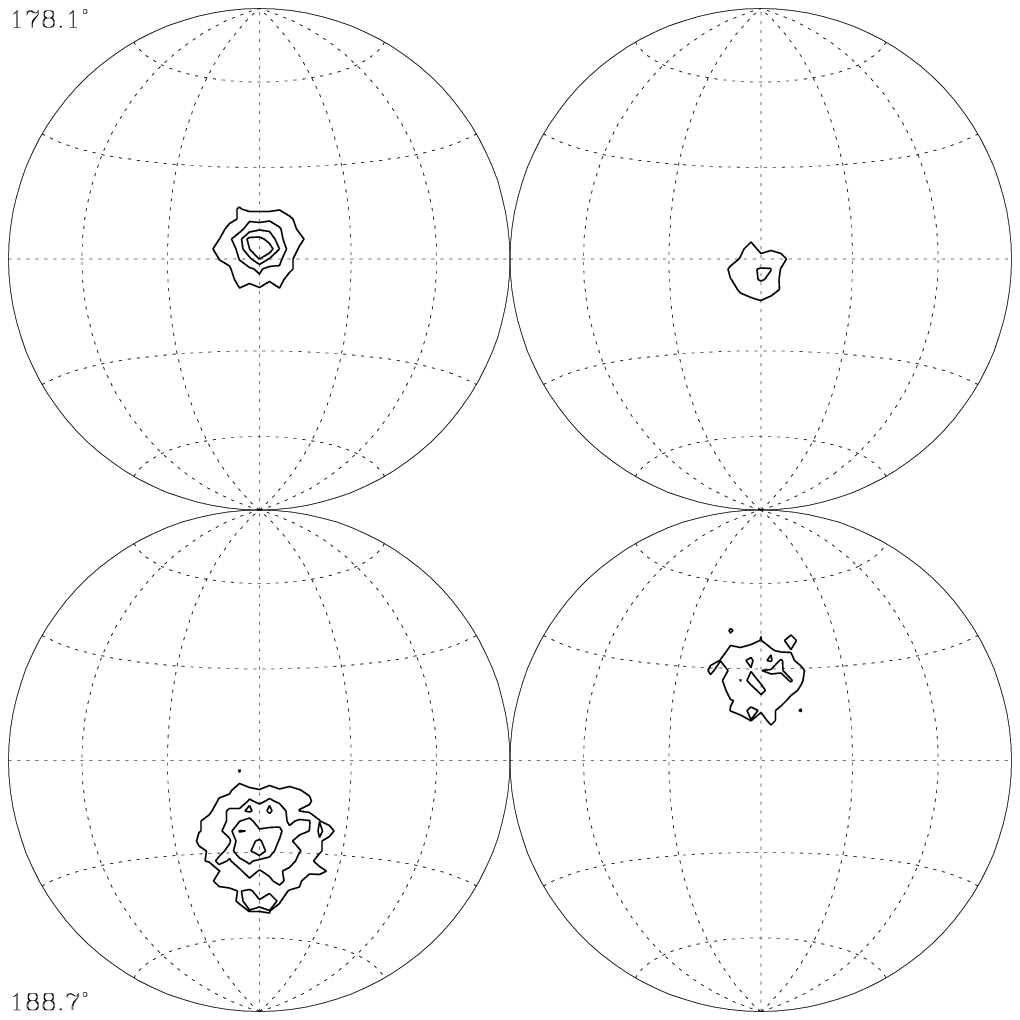}
\caption{Simulations of the PA-EA joint probability density observed in PSR B1451-68. 
Each row shows the Lambert equal-area projection of the two hemispheres of the Poincar\'e 
sphere for a single simulation. The top row shows the results from a simulation of the data 
recorded at a pulse longitude of $178.1^\circ$ (see Figure 2 of Dyks et al. 2021). The 
parameters used in the simulation were $\chi_o=2^\circ$, $s=1.8$, and $\rho=6$. The bottom 
row shows the results from a simulation of the data recorded at a longitude of $188.7^\circ$. 
The parameters used in this simulation were $\chi_o=-13^\circ$, $s=1$, and $\rho=3.8$. The 
contour levels shown in both rows are drawn at 0.2, 0.4, 0.6, and 0.8 of the joint density's
peak value.}
\label{fig:LEAP}
\end{figure}

Figure~\ref{fig:LEAP} shows simulations of the PA-EA joint density at two locations 
within the pulsar's pulse. The OPM statistical model was used in the simulations. Each row 
of the figure is a Lambert equal-area projection of the two hemispheres of the Poincar\'e 
sphere for a single simulation. The simulations were generated by estimating constant values 
of $\chi_o$, $s$, and $\rho$ from the observations and creating 10,000 samples each of the 
Stokes parameters $Q$, $U$, and $V$ from Equations~\ref{eqn:Q}-\ref{eqn:V}. The simulation
modeled the mode intensities as Gaussian random variables. The intrinsic value of the 
dominant mode PA was set to $\psi_o=0$ to simplify the data presentation. Individual samples 
of the PA and EA were calculated from sets of $Q$, $U$, and $V$ and projected on the 
Poincar\'e sphere.\footnote{Note that the Poincar\'e sphere depicts the azimuth, $\phi=2\psi$, 
and latitude, $\lambda=2\chi$, of the polarization vector.} The view of the sphere on the 
left half of the projection is along the $+Q$ axis of the sphere, and the view on the right 
half is along the $-Q$ axis.

The top row of Figure~\ref{fig:LEAP} is a simulation of the PA-EA joint density at a pulse 
longitude of $178.1^\circ$ in PSR B1451-68. This simulation should be compared to the right 
center panel in Figure 2 of Dyks et al. (2021). The observed joint density consists of two
circularly symmetric components. They reside at near-antipodal points close to the equatorial 
plane of the Poincar\'e sphere, indicating that the polarization modes are approximately 
orthogonal and predominantly linearly polarized. One component is more prominent than the 
other. The model parameters estimated from these observed properties and used in the 
simulation were $\chi_o=2^\circ$, $s=1.8$, and $\rho=6$. The results of the simulation are 
qualitatively consistent with what is observed. The simulation also resembles the PA-EA joint 
density observed in the trailing outrider of PSR B0329+54 (see the bottom right corner of 
Figure 2 in Edwards \& Stappers 2004).

The bottom row of the figure is a simulation of the joint density at a pulse longitude of 
$188.7^\circ$. This simulation should be compared to the bottom center panel in Figure 2 
of Dyks et al. (2021). The observed joint density consists of two components residing at 
near-antipodal points on the Poincar\'e sphere, again suggesting that the polarization 
modes are approximately orthogonal. One component is more prominent than the other and is 
concentrated around an EA of $\chi\simeq -13^\circ$, indicating that the emission in this 
region of the pulse is elliptically polarized. The components are broader than those 
observed at a longitude of $178.1^\circ$. The model parameters used in this simulation 
were $\chi_o=-13^\circ$, $s=1$, and $\rho=3.8$. The results of this simulation are also 
qualitatively consistent with what is observed.


\section{Comparison of the Statistical Properties of the Polarization Angles}
\label{sec:compare}

A comparison of the statistical properties of the EA derived in Sections~\ref{sec:fix} 
and~\ref{sec:opm} with those of the PA underscores their similarities and differences. 
When the amplitude of the polarization vector is constant, the pdf of the PA is always 
symmetric about the intrinsic PA, $\psi_o$. Consequently, the mean of the pdf is always 
equal to $\psi_o$. The mean PA is not biased by the instrumental noise, provided the 
noise in the Stokes parameters $Q$ and $U$ is independent and equal in magnitude (Montier 
et al. 2015a, 2015b). The standard deviation of the pdf, $\sigma_\psi$, is independent 
of $\psi_o$ and varies inversely with the SNR in linear polarization. The pdf becomes 
Gaussian as the SNR becomes large, at which point the confidence limits of the pdf are 
well approximated by $\sigma_\psi$. In contrast, the pdf of the EA is generally 
asymmetric. The standard deviation of the pdf can depend upon the value of the intrinsic 
EA, $\chi_o$, and varies inversely with the SNR in total polarization. The measured 
EA is biased by the instrumental noise, particularly at low SNR and large values of 
$\lvert\chi_o\rvert$. Since the pdf is asymmetric, its confidence limits are also 
asymmetric about the mean EA. As the SNR increases, the pdf becomes Gaussian when the 
SNR and intrinsic EA satisfy the condition $s>3/\cos(2\chi_o)$, at which point the mean 
EA is equal to $\chi_o$, $\sigma_\chi$ is independent of $\chi_o$, and the confidence 
limits of the pdf are well approximated by $\sigma_\chi$. The pdf is never symmetric 
when $\chi_o=\pi/4$. 

The pdf of the PA is generally bimodal when the amplitude of the polarization vector
fluctuates due to the superposition of incoherent OPMs. The two components of the pdf 
are separated precisely by $\Delta\psi=\pi/2$. Both components are symmetric about their 
respective centroids, with widths scaling inversely as the ratio of the polarization 
fluctuations relative to the instrumental noise, $\rho$. The pdf of the EA is also bimodal. 
The two components of the pdf are generally asymmetric about their respective peaks, 
and their separation tends to be less than $2\chi_o$, particularly at large values of 
$\lvert\chi_o\rvert$.  As $\rho$ increases, the width of the components decreases, and 
their separation approaches $2\chi_o$. The pdf becomes unimodal when the polarization 
fluctuations are small relative to either the mean polarization or the instrumental noise.


\section{SUMMARY}
\label{sec:summary}

The joint probability density of a polarization vector's orientation angles was derived 
when the vector amplitude is constant or random. The pdf of the EA was derived from the 
joint density. When the vector amplitude is constant, the EA pdf evolves from a 
Gaussian-like function at low intrinsic angles to a mirror image of a Rayleigh 
distribution at an intrinsic EA of $\chi_o=\pi/4$. The measured value of the EA was 
shown to be biased by the instrumental noise, particularly at low values of SNR in 
total polarization, $s$, and large values of $\lvert\chi_o\rvert$. The effect of the 
noise can be reduced with conventional polarization estimation techniques. The standard 
deviation of the EA is generally dependent upon $\chi_o$ and varies inversely with $s$. 
However, when $s$ and $\chi_o$ satisfy the condition $s > 3/\cos(2\chi_o)$, the EA pdf 
is Gaussian, and its standard deviation is independent of $\chi_o$. Since the EA pdf is 
generally asymmetric, its confidence limits tend to be asymmetric about its mean value. 
A method based on the semivariances of the pdf was proposed for calculating its
confidence limits.

The Stokes parameters are covariant when the radio emission from pulsars is comprised 
of superposed incoherent OPMs. In this instance, the joint probability density of the 
polarization angles and the individual pdfs of the EA and PA are unimodal if the mean 
polarization exceeds the polarization fluctuations and are bimodal if the fluctuations 
exceed the mean. When the mean intensities of the modes are equal, the widths of the 
two components in the EA's bimodal pdf vary inversely with the polarization fluctuations 
relative to the instrumental noise. The component separation approaches twice the 
intrinsic EA as the fluctuations increase. 


\appendix

\section{Confidence Limits of the EA}

\begin{figure}
\plotone{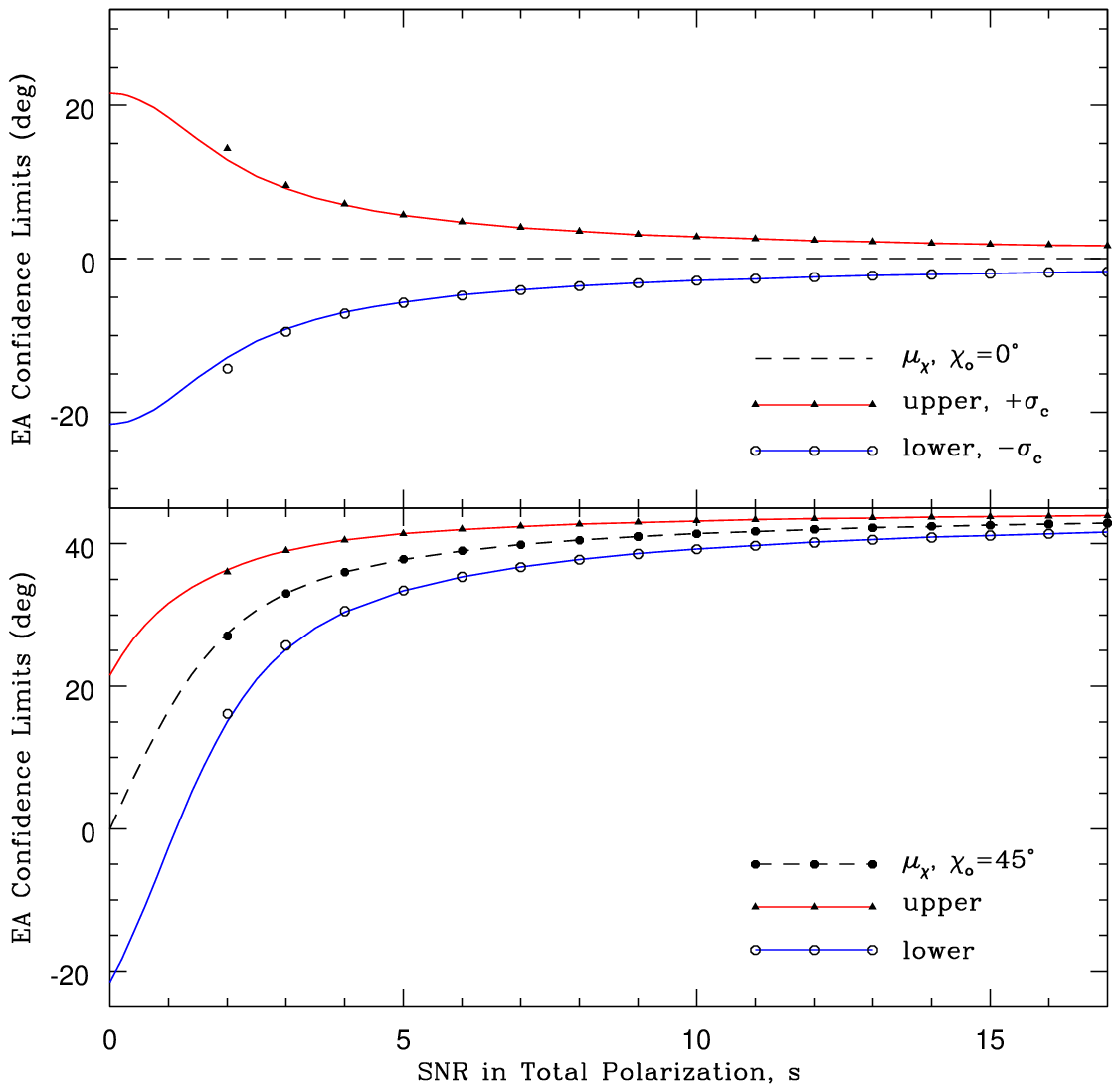}
\caption{Confidence limits ($68\%$) of the EA for $\chi_o=0^\circ$ (top panel) and 
$\chi_o=45^\circ$ (bottom panel) as functions of SNR in total polarization. In each 
panel, the solid red line denotes the upper confidence limit, the solid blue line 
represents the lower confidence limit, and the dashed black line is the mean value of 
the EA calculated from $f_\chi(\chi)$. The symbols attached to each line represent
approximations to the mean value or confidence limits of $\chi$ as discussed in the
text.}
\label{fig:CI}
\end{figure}

The pdf of the EA is symmetric when the amplitude of the polarization vector is 
constant and the intrinsic EA is $\chi_o=0$. The symmetry of the pdf allows a 
$68\%$ confidence limit, $\sigma_c$, to be calculated from Equation~\ref{eqn:chilin},
following the procedure used by Wardle \& Kronberg (1974), Naghizadeh-Khouei \& 
Clarke (1993), and Everett \& Weisberg (2001) to determine the confidence limits 
of the PA:
\begin{equation}
2\int_0^{\sigma_c} f_\chi(\chi)d\chi = 0.6827.
\label{eqn:CI}
\end{equation}
The confidence limit at $s=0$ can be found by inserting $f_\chi(\chi)=\cos(2\chi)$ 
into Equation~\ref{eqn:CI} and is equal to
\begin{equation}
\sigma_c=\frac{1}{2}\arcsin(0.6827) = 21.53^\circ.
\end{equation}
All confidence limits approach this value as the SNR tends to zero, regardless of the 
value of $\chi_o$. The confidence limits calculated from Equation~\ref{eqn:CI} are shown in 
the top panel of Figure~\ref{fig:CI}. The limits are symmetric about the mean because the 
pdf is symmetric. The filled triangles and open circles in the panel denote the approximation 
to the standard deviation of the pdf given by Equation~\ref{eqn:chiSD0}. The approximation 
is a reasonable facsimile of the confidence limits down to an SNR of $s\simeq 3$.

A different approach to calculating the confidence limits must be adopted for other values
of $\chi_o$, because $f_\chi(\chi)$ is generally asymmetric in those cases. Przysucha (2024) 
suggested that the confidence interval of an asymmetric pdf, $f(x)$, can be calculated from
its semivariances. His method is applicable to any continuous random variable with a mean, 
$\mu$, and standard deviation, $\sigma$. The upper semivariance, $\sigma_+^2$, is the 
variance of $f(x)$ calculated from values of $x$ that exceed the mean:
\begin{equation}
\sigma_+^2 = \int_{\mu}^{\infty}(x-\mu)^2f(x)dx;
\end{equation}
and the lower semivariance, $\sigma_-^2$, is the variance of $f(x)$ calculated
from values of $x$ that fall below the mean:
\begin{equation}
\sigma_-^2 = \int_{-\infty}^{\mu}(x-\mu)^2f(x)dx.
\end{equation}
The sum of the semivariances is always equal to the variance of $f(x)$, 
$\sigma^2 = \sigma_+^2 + \sigma_-^2$. When $f(x)$ is symmetric, the semivariances are 
equal to one-half the variance, $\sigma_+^2 = \sigma_-^2 = \sigma^2/2$. 

Przysucha's (2024) confidence interval, $\gamma$, is given by
\begin{equation}
\gamma = P\{-k\sigma_-\le x - \mu \le k\sigma_+\}.
\label{eqn:pci}
\end{equation}
Increasing the parameter $k$ in Equation~\ref{eqn:pci} increases the confidence 
interval and the confidence limits. The confidence limits of the EA can be 
determined by selecting a desired value of the confidence interval - in this case 
$\gamma=0.6827$ - and finding the value of $k$ that is the solution to
\begin{equation}
\int_{\tau_-}^{\tau_+}f_\chi(\chi)d\chi = 0.6827.
\label{eqn:chici}
\end{equation}
The lower limit of integration in Equation~\ref{eqn:chici} is $\tau_-=\mu_\chi-k\sigma_-$ 
and the upper limit is $\tau_+=\mu_\chi+k\sigma_+$. Once $k$ is known, the confidence limits 
can be calculated from the definitions of the semivariances. The value of $k$ varies with 
different values of $s$ and $\chi_o$ in $f_\chi(\chi)$. It is equal to $\sqrt{2}$ when the 
pdf is Gaussian, because the semivariances are equal to one-half the pdf's variance, and 
the $68\%$ confidence limit of a Gaussian distribution is equal to its standard deviation.

The resulting upper and lower confidence limits in the case of $\chi_0=\pi/4$ are shown 
by the red and blue lines, respectively, in the bottom panel of Figure~\ref{fig:CI}. The 
dashed black line denotes the mean EA, as calculated from the first moment of the pdf given 
by Equation~\ref{eqn:pdf45}. The filled circles that accompany the dashed black line denote 
the approximation to the mean EA given by Equation~\ref{eqn:avg45}. The approximation is 
very good. The figure shows the confidence limits generally are not symmetric about the 
mean. A fit to the confidence limits calculated from Equation~\ref{eqn:chici} yields an 
upper bound on $\chi$ that varies with $s$ as
\begin{equation}
\tau_+ = 45.0^\circ-\frac{17.97^\circ}{s},
\label{eqn:hi45}
\end{equation}
and a lower bound that varies as
\begin{equation}
\tau_- = 45.0^\circ-\frac{57.78^\circ}{s}.
\label{eqn:lo45}
\end{equation}
These approximations are shown by the filled triangles and open circles, respectively, in the 
panel. A comparison of Equations~\ref{eqn:hi45} and~\ref{eqn:lo45} with Equation~\ref{eqn:mode45} 
shows that the mode of the EA always resides within the $68\%$ confidence interval.

\def\arraystretch{1.25} 

\begin{deluxetable}{crrrrrr}
\tablenum{1}
\tablecaption{Mean and 68\% Confidence Limits of the EA}
\tablehead{\colhead{$\chi_o(^\circ)$} & \colhead{$s=3$} & \colhead{$s=6$} & \colhead{$s=9$} 
  & \colhead{$s=12$} & \colhead{$s=15$} & \colhead{$s=18$}}
\startdata
$0$ & $0.0_{-9.2}^{+9.2}$ & $0.0_{-4.7}^{+4.7}$ & $0.0_{-3.2}^{+3.2}$ & $0.0_{-2.4}^{+2.4}$ &
  $0.0_{-1.9}^{+1.9}$ & $0.0_{-1.6}^{+1.6}$ \\
$5$ & $4.7_{-9.2}^{+9.1}$ & $4.9_{-4.7}^{+4.7}$ & $5.0_{-3.2}^{+3.2}$ & $5.0_{-2.4}^{+2.4}$ & 
  $5.0_{-1.9}^{+1.9}$ & $5.0_{-1.6}^{+1.6}$ \\
$10$ & $9.3_{-9.2}^{+9.0}$ & $9.8_{-4.7}^{+4.7}$ & $9.9_{-3.2}^{+3.2}$ & $10.0_{-2.4}^{+2.4}$ & 
  $10.0_{-1.9}^{+1.9}$ & $10.0_{-1.6}^{+1.6}$ \\
$15$ & $14.0_{-9.2}^{+8.9}$ & $14.8_{-4.7}^{+4.7}$ & $14.9_{-3.2}^{+3.2}$ & $14.9_{-2.4}^{+2.4}$ & 
  $15.0_{-1.9}^{+1.9}$ & $15.0_{-1.6}^{+1.6}$ \\
$20$ & $18.4_{-9.2}^{+8.7}$ & $19.7_{-4.7}^{+4.7}$ & $19.8_{-3.2}^{+3.2}$ & $19.9_{-2.4}^{+2.4}$ & 
  $20.0_{-1.9}^{+1.9}$ & $20.0_{-1.6}^{+1.6}$ \\
$25$ & $22.8_{-9.2}^{+8.4}$ & $24.5_{-4.7}^{+4.7}$ & $24.8_{-3.2}^{+3.2}$ & $24.9_{-2.4}^{+2.4}$ & 
  $24.9_{-1.9}^{+1.9}$ & $25.0_{-1.6}^{+1.6}$ \\
$30$ & $26.7_{-9.1}^{+7.9}$ & $29.3_{-4.7}^{+4.6}$ & $29.7_{-3.2}^{+3.1}$ & $29.8_{-2.4}^{+2.4}$ &
  $29.9_{-1.9}^{+1.9}$ & $29.9_{-1.6}^{+1.6}$ \\
$35$ & $30.0_{-8.7}^{+7.1}$ & $33.8_{-4.6}^{+4.4}$ & $34.5_{-3.1}^{+3.1}$ & $34.7_{-2.4}^{+2.3}$ & 
  $34.8_{-1.9}^{+1.9}$ & $34.9_{-1.6}^{+1.6}$ \\
$40$ & $32.2_{-8.2}^{+6.3}$ & $37.5_{-4.2}^{+3.6}$ & $38.9_{-3.0}^{+2.7}$ & $39.4_{-2.3}^{+2.2}$ & 
  $39.6_{-1.9}^{+1.8}$ & $39.7_{-1.6}^{+1.5}$ \\
$45$ & $33.0_{-7.9}^{+6.0}$ & $39.0_{-3.6}^{+3.0}$ & $41.0_{-2.4}^{+2.0}$ & $42.0_{-1.8}^{+1.5}$ &
  $42.6_{-1.4}^{+1.2}$ & $43.0_{-1.2}^{+1.0}$ \\
\enddata
\end{deluxetable}

\def\arraystretch{1.0} 

Example confidence limits calculated for other values of $\chi_o$ are tabulated in Table 1. The 
left column of the table lists the intrinsic EA, ranging from $0^\circ$ to $45^\circ$ in increments
of $\Delta\chi_o=5^\circ$. The top row of the table lists values of the SNR in total polarization,
ranging from $s=3$ to $s=18$ in increments of $\Delta s=3$. The table entries include the means of 
the pdfs defined by the specified values of $s$ and $\chi_o$, along with the upper and lower errors 
determined from Equation~\ref{eqn:chici}. The magnitudes of the upper and lower errors for a given 
table entry are the same for most entries, especially when the SNR is large and the intrinsic EA 
is small. This is an indication that their respective pdfs are symmetric, and their values of $s$ 
and $\chi_o$ are consistent with the condition $s>3/\cos(2\chi_o)$. The magnitude of the errors 
for the individual entries in the second column and last two rows of the table generally are not 
equal, indicating their respective pdfs are not symmetric. 

Table 1 can serve as a lookup table for placing upper and lower limits on an EA measurement.
Assuming that the measured EA is the mean EA, the table entries and the measured values of $s$ 
provide the errors associated with the measurement. For example, if the measured SNR is $s=3$ 
and the measured EA is $\chi_m=26.7^\circ$, the upper error on the measurement is $+7.9^\circ$, 
and the lower error is $-9.1^\circ$. Alternatively, analytical expressions for the upper and 
lower bounds as functions of $s$ for a given $\chi_o$ can be derived by fitting the data in 
Table 1 to equations similar to Equations~\ref{eqn:hi45} and~\ref{eqn:lo45}. Or a more
extensive lookup table with finer resolution in $s$ and $\chi_o$ can be generated, by 
numerically calculating the mean of the relevant pdf and its upper and lower bounds from 
Equation~\ref{eqn:chici}.



\begin{acknowledgments}

I thank the anonymous reviewers for constructive comments on the original 
version of the manuscript. I also thank Dan Stinebring for providing the data used to produce 
Figure~\ref{fig:eahist}. The National Radio Astronomy Observatory and Green Bank Observatory 
are facilities of the U.S. National Science Foundation operated under cooperative agreement 
by Associated Universities, Inc.

\end{acknowledgments}



\begin{references}
\reference{} Allen, M. C. \& Melrose, D. B. 1982, PASA, 4, 365, 
             doi:10.1017/S1323358000021147

\reference{} Backer, D. C. \& Rankin, J. M. 1980, \apjs, 42, 143,
             doi: 10.1086/190647

\reference{} Barnard, J. J. \& Arons, J. A., 1986, \apj, 302, 138,
             doi: 10.1086/163979

\reference{} Bera, A., James, C., McKinnon, M. M., et al., 2025, \apj, 982, 119,
             doi: 10.3847/1538-4357/adba59

\reference{} Cao, S., Jiang, J., Dyks, J., et al., 2025, \apj, 983, 43,
             doi: 10.3847/1538-4357/adbe33

\reference{} Cordes, J. M., Rankin, J., \& Backer, D. C. 1978, \apj, 223, 961,
             doi: 10.1086/156328

\reference{} Cordes, J. M. 1981, in IAU Symp. 95, Pulsars, ed. W. Sieber \& R.
             Wielebinski (Dordrecht:Reidel), 115

\reference{} Davenport, W. B. \& Root, W. L. 1958, Introduction to the Theory of
             Random Signals and Noise (New York: McGraw-Hill)

\reference{} Dyks, J., Weltevrede, P., \& Ilie, C. 2021, \mnras, 501, 2156,
             doi: 10.1093/mnras/staa3762

\reference{} Edwards, R. T. 2004, \aap, 426, 677, doi: 10.1051/0004-6361:20041029

\reference{} Edwards, R. T. \& Stappers, B. W. 2004, \aap, 421, 681,
             doi: 10.1051/0004-6361:20040228

\reference{} Everett, J. E. \& Weisberg, J. M. 2001, \apj, 553, 341,
             doi: 10.1086/320652

\reference{} Ilie, C. D., Weltevrede, P., Johnston, S. \& Chen, T. 2020, \mnras, 491, 3385,
             doi: 10.1093/mnras/stz3167

\reference{} Kennett, M. \& Melrose, D. 1998, PASA, 15, 211,
             doi: 10.1071/AS98211

\reference{} Lower, M. E., Johnston, S., Lyutikov, M., et al., 2024, NatAs, 8, 606,
             doi: 10.1038/s41550-024-02225-8

\reference{} Manchester, R. N., Taylor, J. H., \& Huguenin, G. R. 1975, \apj, 196, 83,
             doi: 10.1086/153395

\reference{} McKinnon, M. M. 2003, \apjs, 148, 519, doi: 10.1086/376898

\reference{} McKinnon, M. M. 2004, \apj, 606, 1154, doi: 10.1086/383194

\reference{} McKinnon, M. M. 2006, \apj, 645, 551, doi: 10.1086/504314 

\reference{} McKinnon, M. M. 2022, \apj, 937, 92, doi: 10.3847/1538-4357/ac8dfa

\reference{} McKinnon, M. M. \& Stinebring, D. R. 1998, \apj, 502, 883,
             doi: 10.1086/305924

\reference{} McKinnon, M. M. \& Stinebring, D. R. 2000, \apj, 529, 435,
             doi: 10.1086/308264

\reference{} Melrose, D. B., 1979, AuJPh, 32, 61, doi: 10.1071/PH790061

\reference{} Melrose, D. B. \& Stoneham, R. J., 1977, PASA, 3, 120, 
             doi: 10.1017/S1323358000015010

\reference{} Montier, L., Plaszczynski, S., Levrier, F. et al., 2015a, \aap, 574, A135,
             doi: 10.1051/0004-6361/201322271

\reference{} Montier, L., Plaszczynski, S., Levrier, F. et al., 2015b, \aap, 574, A136,
             doi: 10.1051/0004-6361/201424451

\reference{} Naghizadeh-Khouei, J. \& Clarke, D. 1993, \aap, 274, 968

\reference{} Oswald, L. S., Karastergiou, A., \& Johnston, S. 2023, \mnras, 525, 840,
             doi:10.1093/mnras/stad2271

\reference{} Papoulis, A., 1965, Probability, Random Variables, and Stochastic Processes
             (New York: McGraw-Hill)

\reference{} Plaszczynski, S., Montier, L., Levrier, F. et al. 2014, \mnras, 439, 4048,
             doi: 10.1093/mnras/stu270

\reference{} Przysucha, B., 2024, Metrol. Meas. Syst., 31, 279, 
             doi: 10.24425/mms.2024.149701

\reference{} Serkowski, K. 1958, AcA, 8, 135

\reference{} Simmons, J. F. L. \& Stewart, B.G., 1985, \aap, 142, 100

\reference{} Stinebring, D. R., Cordes, J. M., Rankin, J. M., et al. 1984, \apjs, 55, 247,
             doi: 10.1086/190954

\reference{} Wardle, J. F. C. \& Kronberg, P. P., 1974, \apj, 194, 249,
             doi: 10.1086/153240

\end{references}
\end{document}